\documentclass[twocolumn]{aastex631}
\pdfoutput=1 
\usepackage[T1]{fontenc}
\usepackage{apjfonts} 
\usepackage[figure,figure*]{hypcap}
\usepackage{comment}
\usepackage{amssymb}
\usepackage{amsmath}
\usepackage[T1]{tipa}
\usepackage{CJKutf8}

\newcommand{\frb}{FRB~20240209A}
\newcommand{\kkoname}{k'ni\textipa{P}atn k'l$\left._\mathrm{\smile}\right.$stk'masqt}
\newcommand{\kkonamecaps}{K'ni\textipa{P}atn k'l$\left._\mathrm{\smile}\right.$stk'masqt}
\begin{document}

\shortauthors{Shah et al.}

\shorttitle{A repeating FRB in the outskirts of a quiescent galaxy}

\title{A repeating fast radio burst source in the outskirts of a quiescent galaxy}

\newcommand{\ICRAR}{\affiliation{International Centre for Radio Astronomy Research (ICRAR), Curtin University, Bentley WA 6102 Australia}}

\newcommand{\DRAO}{\affiliation{Dominion Radio Astrophysical Observatory, Herzberg Research Centre for Astronomy and Astrophysics, National Research Council Canada, PO Box 248, Penticton, BC V2A 6J9, Canada}}

\newcommand{\CSIRO}{\affiliation{CSIRO Space \& Astronomy, Parkes Observatory, P.O. Box 276, Parkes NSW 2870, Australia}}

\newcommand{\NRC}{\affiliation{NRC Herzberg Astronomy and Astrophysics, 5071 West Saanich Road, Victoria, BC V9E2E7, Canada}}

\newcommand{\CIERA}{\affiliation{Center for Interdisciplinary Research in Astronomy, Northwestern University, 1800 Sherman Avenue, Evanston, IL 60201, USA }}

\newcommand{\NU}{\affiliation{Department of Physics and Astronomy, Northwestern University, Evanston, IL 60208, USA}}

\newcommand{\Uch}
{\affiliation{Department of Astronomy and Astrophysics, University of Chicago, William Eckhart Research Center, 5640 South Ellis Avenue, Chicago, IL 60637, USA}}

\newcommand{\UCSC}{\affiliation{Department of Astronomy and Astrophysics, University of California Santa Cruz, 1156 High Street, Santa Cruz, CA 95064, USA}}

\newcommand{\IPMU}{\affiliation{Kavli Institute for the Physics and Mathematics of the Universe (Kavli IPMU), 5-1-5 Kashiwanoha, Kashiwa, 277-8583, Japan}}

\newcommand{\NAOJ}{\affiliation{Division of Science, National Astronomical Observatory of Japan, 2-21-1 Osawa, Mitaka, Tokyo 181-8588, Japan}}

\newcommand{\MU}{\affiliation{Department of Physics, McGill University, 3600 rue University, Montr\'eal, QC H3A 2T8, Canada}}

\newcommand{\Trottier}{\affiliation{Trottier Space Institute, McGill University, 3550 rue University, Montr\'eal, QC H3A 2A7, Canada}}

\newcommand{\CMU}{\affiliation{McWilliams Center for Cosmology \& Astrophysics, Department of Physics, Carnegie Mellon University, Pittsburgh, PA 15213, USA}}

\newcommand{\UVA}
{\affiliation{Anton Pannekoek Institute for Astronomy, University of Amsterdam, Science Park 904, 1098 XH Amsterdam, The Netherlands}}

\newcommand{\ASTRON}
{\affiliation{ASTRON, Netherlands Institute for Radio Astronomy, Oude Hoogeveensedijk 4, 7991 PD Dwingeloo, The Netherlands
}}

\newcommand{\MITK}
{\affiliation{MIT Kavli Institute for Astrophysics and Space Research, Massachusetts Institute of Technology, 77 Massachusetts Ave, Cambridge, MA 02139, USA}}

\newcommand{\MITP}
{\affiliation{Department of Physics, Massachusetts Institute of Technology, 77 Massachusetts Ave, Cambridge, MA 02139, USA}}

\newcommand{\CCAPS}{\affiliation{Cornell Center for Astrophysics and Planetary Science, Cornell University, Ithaca, NY 14853, USA}}

\newcommand{\DI}
{\affiliation{Dunlap Institute for Astronomy and Astrophysics, 50 St. George Street, University of Toronto, ON M5S 3H4, Canada}}

\newcommand{\DAA}
{\affiliation{David A. Dunlap Department of Astronomy and Astrophysics, 50 St. George Street, University of Toronto, ON M5S 3H4, Canada}}

\newcommand{\STSCI}
{\affiliation{Space Telescope Science Institute, 3700 San Martin Drive, Baltimore, MD 21218, USA}}

\newcommand{\WVUPHAS}
{\affiliation{Department of Physics and Astronomy, West Virginia University, PO Box 6315, Morgantown, WV 26506, USA }}

\newcommand{\WVUGWAC}
{\affiliation{Center for Gravitational Waves and Cosmology, West Virginia University, Chestnut Ridge Research Building, Morgantown, WV 26505, USA}}

\newcommand{\UCB}
{\affiliation{Department of Astronomy, University of California, Berkeley, CA 94720, United States}}

\newcommand{\YORK}
{\affiliation{Department of Physics and Astronomy, York University, 4700 Keele Street, Toronto, ON MJ3 1P3, Canada}}

\newcommand{\PI}
{\affiliation{Perimeter Institute of Theoretical Physics, 31 Caroline Street North, Waterloo, ON N2L 2Y5, Canada}}

\newcommand{\UBC}
{\affiliation{Department of Physics and Astronomy, University of British Columbia, 6224 Agricultural Road, Vancouver, BC V6T 1Z1 Canada}}

\newcommand{\UCHILE}
{\affiliation{Department of Electrical Engineering, Universidad de Chile, Av. Tupper 2007, Santiago 8370451, Chile}}

\author[0000-0002-4823-1946]{V.~Shah}
\MU
\Trottier

\author[0000-0002-6823-2073]{K.~Shin}
\MITK
\MITP

\author[0000-0002-4209-7408]{C. Leung}
\altaffiliation{NHFP Einstein Fellow}
\UCB

\author[0000-0002-7374-935X]{W.~Fong}
\CIERA
\NU

\author[0000-0003-0307-9984]{T.~Eftekhari}
\altaffiliation{NHFP Einstein Fellow}
\CIERA
%


\author[0000-0001-6523-9029]{M.~Amiri}
\UBC

\author[0000-0001-5908-3152]{B.~C.~Andersen}
\MU
\Trottier

\author[0000-0002-3980-815X]{S.~Andrew}
\MITK
\MITP

\author[0000-0002-3615-3514]{M.~Bhardwaj}
\CMU

\author[0000-0002-1800-8233]{C.~Brar}
\NRC

\author[0000-0003-2047-5276]{T.~Cassanelli}
\UCHILE

\author[0000-0002-2878-1502]{S.~Chatterjee}
\CCAPS

\author[0000-0002-8376-1563]{A.~P.~Curtin}
\MU
\Trottier

\author[0000-0001-7166-6422]{M.~Dobbs}
\MU
\Trottier

\author[0000-0002-9363-8606]{Y.~Dong \begin{CJK*}{UTF8}{gbsn}(董雨欣)\end{CJK*}}
\CIERA
\NU

\author[0000-0003-4098-5222]{F.~A.~Dong}
\UBC

\author[0000-0001-8384-5049]{E.~Fonseca}
\WVUPHAS
\WVUGWAC

\author[0000-0002-3382-9558]{B.~M.~Gaensler}
\UCSC
\DAA
\DI

\author[0000-0002-1760-0868]{M.~Halpern}
\UBC

\author[0000-0003-2317-1446]{J.~W.~T.~Hessels}
\MU
\Trottier
\UVA
\ASTRON

\author[0000-0003-2405-2967]{A.~L.~Ibik}
\DI
\DAA

\author[0009-0009-0938-1595]{N.~Jain}
\MU
\Trottier

\author[0000-0003-3457-4670]{R.~C.~Joseph}
\MU
\Trottier

\author[0000-0003-4810-7803]{J.~Kaczmarek}
\CSIRO

\author[0009-0007-5296-4046]{L.~Kahinga}
\UCSC

\author[0000-0001-9345-0307]{V.~Kaspi}
\MU 
\Trottier

\author[0009-0008-6166-1095]{B.~Kharel}
\WVUPHAS
\WVUGWAC

\author[0000-0003-1455-2546]{T.~Landecker}
\DRAO

\author[0000-0003-2116-3573]{A.~E.~Lanman}
\MITK
\MITP

\author[0000-0002-5857-4264]{M.~Lazda}
\DAA
\DI

\author[0000-0002-7164-9507]{R.~Main}
\MU

\author[0000-0003-4584-8841]{L.~Mas-Ribas}
\UCSC

\author[0000-0002-4279-6946]{K.~W.~Masui}
\MITK
\MITP

\author[0000-0001-7348-6900]{R.~Mckinven}
\MU
\Trottier

\author[0000-0002-0772-9326]{J.~Mena-Parra}
\DAA
\DI

\author[0000-0001-8845-1225]{B.~W.~Meyers}
\ICRAR

\author[0000-0002-2551-7554]{D.~Michilli}
\MITK
\MITP

\author[0000-0003-0510-0740]{K.~Nimmo}
\MITK

\author[0000-0002-8897-1973]{A.~Pandhi}
\DAA
\DI

\author[0009-0008-7264-1778]{S.~S.~Patil}
\WVUPHAS
\WVUGWAC

\author[0000-0002-8912-0732]{A.~B.~Pearlman}
\altaffiliation{Banting Fellow, McGill Space Institute~(MSI) Fellow, \\ and FRQNT Postdoctoral Fellow.}
\MU
\Trottier

\author[0000-0002-4795-697X]{Z.~Pleunis}
\UVA
\ASTRON

\author[0000-0002-7738-6875]{J.~X.~Prochaska}
\UCSC
\IPMU
\NAOJ

\author[0000-0001-7694-6650]{M.~Rafiei-Ravandi}
\MU

\author[0000-0002-4623-5329]{M.~Sammons}
\MU
\Trottier

\author[0000-0003-3154-3676]{K.~R.~Sand}
\MU
\Trottier

\author[0000-0002-7374-7119]{P.~Scholz}
\YORK
\DI 

\author[0000-0002-2088-3125]{K.~Smith}
\PI

\author[0000-0001-9784-8670]{I.~Stairs}
\UBC

\correspondingauthor{Vishwangi Shah}
\email{vishwangi.shah@mail.mcgill.ca}

\begin{abstract}
    We report the discovery of the repeating fast radio burst source \frb{} using the CHIME/FRB telescope. We have detected 22 bursts from this repeater between February and July 2024, six of which were also recorded at the Outrigger station \kkoname{} (KKO). The 66-km long CHIME--KKO baseline can provide single-pulse FRB localizations along one dimension with $2^{\prime\prime}$ accuracy. The high declination of $\sim$86 degrees for this repeater allowed its detection with a rotating range of baseline vectors, enabling the combined localization region size to be constrained to $1^{\prime\prime}\times2^{\prime\prime}$. We present deep Gemini observations that, combined with the FRB localization, enabled a robust association of \frb{} to the outskirts of a luminous 
    galaxy (P(O|x) = 0.99; $L \approx 5.3 \times 10^{10}\,L_{\odot}$). \frb{} has a projected physical offset of $40 \pm 5$\,kpc from the center of its host galaxy, making it the FRB with the largest host galaxy offset to date. When normalized by the host galaxy size, the offset of \frb{} is comparable to that of FRB~20200120E, the only FRB source known to originate in a globular cluster. We consider several explanations for the large offset, including a progenitor that was kicked from the host galaxy or {\it in situ} formation in a low-luminosity satellite galaxy of the putative host, but find the most plausible scenario to be a globular cluster origin. This, coupled with the quiescent, elliptical nature of the host as demonstrated in our companion paper, provide strong evidence for a delayed formation channel for the progenitor of the FRB source. 
\end{abstract} 

\keywords{Radio bursts (1339) --- Radio transient sources (2008) --- Very long baseline interferometry (1769) --- Galaxies(573)}

\section{Introduction} \label{sec:intro}

Fast radio bursts (FRBs) are $\sim$micro-to-millisecond duration bursts of radio emission originating from extragalactic distances \citep{Petroff_FRB_overview}. While thousands of FRBs have been detected, their origins remain unknown. Most FRB progenitor models involve stellar populations such as neutron stars and magnetars \citep{Wenbin2018_magnetar, Metzger2019_magnetar, Wenbin2020_magnetar, Bochenek2021_magnetar}. Some FRBs repeat, which rules out cataclysmic progenitor models for at least this class of FRBs. Repeating FRBs have varying burst activity rates, with some having sudden periods of heightened activity \citep{Adam_R67, Kenzie_M81_burst_storm, FAST_R117}, while one repeater has bursts clustered in periodic activity windows \citep{R3_periodicity}. The detection of repeat FRB-like bursts from a Galactic magnetar \citep{Galactic_Magnetar, Bochenek_sgr_1935} makes a strong case for magnetars as the origin of at least some FRBs.  

Localizing FRBs to their local environments and host galaxies is one of the primary ways to uncover their origins. One-off FRBs have been localized to their host galaxies via connected-element interferometry \citep{ASAKP_host_gal, Meerkat_host_gal, DSA_host_gal_catalog}, and follow-up of repeating FRBs has enabled their localization to their local environments via very-long-baseline interferometry \citep[VLBI;][]{121102_loc_EVN, Marcote_2020, EVN_M81R, Nimmo_2022, Hewitt_2024, CHIME_VLBI_pathfinders}. There is some indication that FRBs preferentially occur in star-forming galaxies, with only a handful of FRBs localized to quiescent environments \citep{Sharma2024}. Additionally, non-repeating FRBs are claimed to predominantly have spiral host galaxies \citep{Mannings_spiral_gal, Mohit_spiral_galaxies}, with only one FRB having a candidate elliptical host \citep{Sharma2024}. Repeating FRBs have been localized to a variety of local environments, from extreme magneto-ionic environments within dwarf star-forming galaxies \citep{121102_loc_VLA, Danille_R1_RM, R1_twin_PRS, R1_twin_RM} to a globular cluster of the massive spiral galaxy M81 \citep{EVN_M81R}. These varying local environments and host galaxies indicate that FRBs can occur in regions of both high and low star formation, and thus both prompt and delayed formation channels for FRB progenitors are possible. 

The Canadian Hydrogen Intensity Mapping Experiment Fast Radio Burst (CHIME/FRB) project \citep{CHIME/FRB_overview} has detected the largest number of FRBs, both repeating and non-repeating \citep{CHIME_catalog_1, Emmanuel_nine_repeaters, RN3}. Some of these FRBs have channelized raw voltage data (hereafter referred to as baseband data) saved; however, interferometric localizations using the baseband data from CHIME alone can only provide an angular precision of $\sim1^{\prime}$ \citep{baseband_loc}, which is not precise enough to associate most FRBs to their host galaxies \citep{FRB_host_gal}. Consequently, CHIME/FRB is building three Outrigger telescopes to form a VLBI array with CHIME to be able to localize FRBs with sub-arcsecond precision \citep{leung2021synoptic, JMP_clk, TONE, CHIME_VLBI_pathfinders, KKO_overview, Aaron_TB, PyFX, Shion_cal_survey}.

Here we report, after $\sim$1500 hours of CHIME/FRB exposure over a period of 6 years, the discovery of an active repeater \frb{}. Using VLBI between CHIME and one of the outrigger stations \kkoname{}\footnote{From the upper Similkameen language, this translates to ``a listening device for outer space''.} \citep[KKO;][]{KKO_overview}, we have localized \frb{} to $\sim$arcsecond precision. In Section~\ref{sec:observations}, we describe the observations of this source and provide its burst rate estimates. In Section~\ref{sec:localization}, we present the CHIME--KKO localization. The identification and nature of the host galaxy for this repeater are discussed in Section~\ref{sec:hostgalaxy}. In Section~\ref{sec:discussion}, we compare the properties of this FRB source to other repeating FRBs and make predictions about the possible progenitors based on the localization region and host galaxy. A detailed discussion about the host galaxy properties is presented by \citet{Eftekhari_R155}. 

\begin{figure*}[t]
\centering
\includegraphics[width = 1\textwidth]{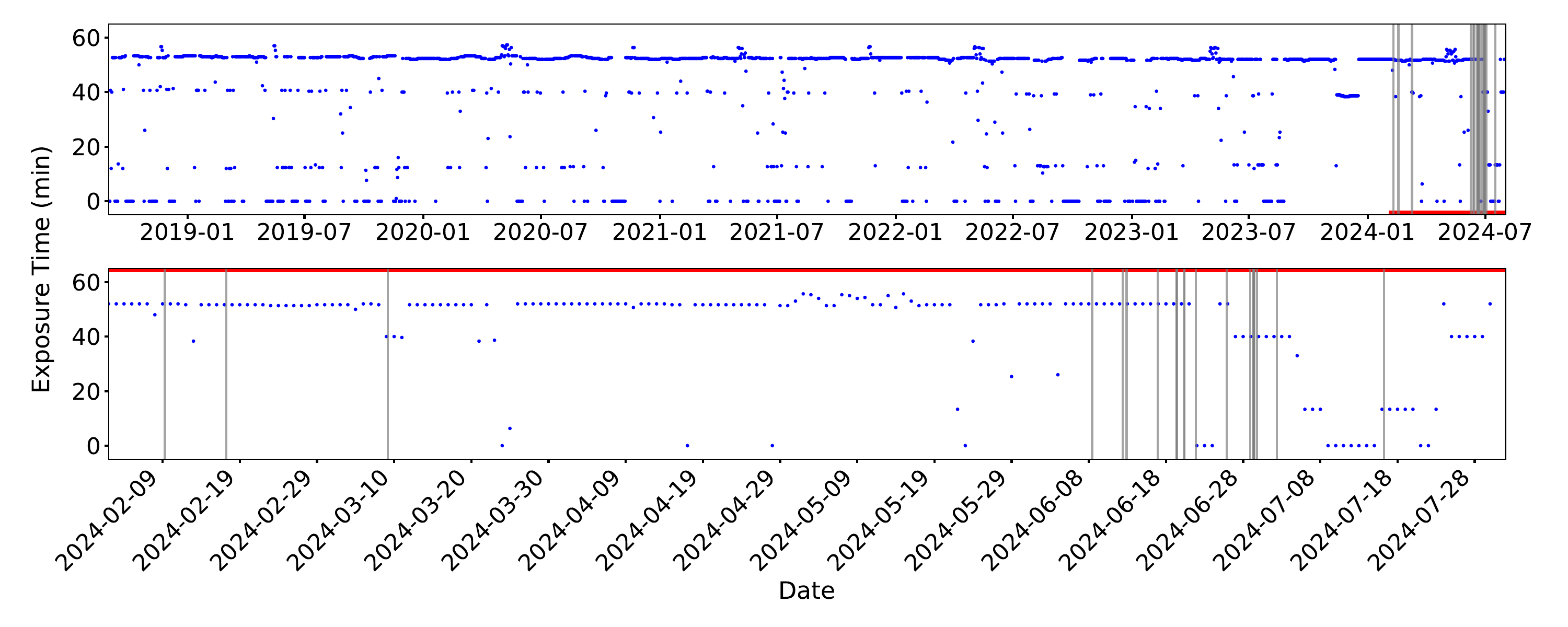}
\caption{Total daily exposure time (UTC and topocentric at CHIME near Penticton, Canada) at the position of \frb{} (blue points) and detection times of the repeat bursts (gray vertical lines). The top panel shows the exposure times since CHIME/FRB began operations. The red line indicates a window around the time the repeater became active, and the bottom panel is zoomed in on that window. The source had a median total exposure of approximately 55\,mins per day, which includes upper and lower transits having median exposures of 13\,mins and 39\,mins, respectively. Days with lower exposure are due to system shutdowns. The exposure is slightly higher on certain days every year due to an additional fractional sidereal transit of the source on the same solar day.}\label{fig:exposure}
\end{figure*}

\section{Observations \label{sec:observations}}
CHIME is a transit radio telescope located at the Dominion Radio Astrophysical Observatory (DRAO) near Penticton, British Columbia. It has a large field of view of $>$ 200\,deg$^2$, a wide frequency bandwidth of $400-800$\,MHz, and records data in the East-West (E-W) and North-South (N-S) linear polarization bases \citep{CHIME_overview}. The CHIME/FRB project uses the CHIME instrument to continuously scan the entire Northern sky and search for FRBs in a multistage process described by \cite{CHIME/FRB_overview}. CHIME/FRB records intensity (Stokes-I) data with 0.98\,ms time resolution for all candidate FRB events with S/N $>$ 8. It also records baseband data with a 2.56\,$\mu$s time resolution for events with S/N $>$ 12 and for all events identified as candidate repeaters with S/N $>$ 10. 
 
The CHIME/FRB Outrigger station KKO provides a 66-km long East-West baseline with CHIME. The statistical localization precision achievable for this baseline for an event of S/N = 12 and having a bandwidth of 400\,MHz is $\sim1^{\prime\prime}$ \citep{Rogers_VLBI}. However, the realistic CHIME--KKO localization accuracy, which also includes systematic errors, is $\sim2^{\prime\prime}$ along the baseline vector \citep{KKO_overview}. KKO has 1/16 the collecting area of CHIME and is very similar to CHIME in terms of its optical design, as well as its analog and digital systems. It is tilted by $\sim0.5^{\circ}$ from the zenith in order to see the same sky as CHIME. KKO does not have an FRB detection backend, but it continuously buffers baseband data in a rolling buffer. CHIME/FRB sends FRB triggers to KKO for bursts with S/N $>$ 15, upon which $\sim$100\,ms of the buffered data around the time of the FRB are saved to disk at KKO such that the data in each frequency channel follows the dispersive sweep of the FRB. Note that in this paper we use CHIME/FRB to refer to the central telescope system that searches for FRBs in real-time, while we use KKO to refer to the Outrigger station that can trigger baseband data recording when a burst is detected by the central system.

\frb{} was first detected on 2024 February 09 at 07:10:14 UTC (topocentric at CHIME near Penticton, Canada). Since then, 21 repeat bursts with real-time positions and dispersion measures (DMs) consistent with the first burst have been detected, until 2024 July 31. Fifteen of these bursts are reported by \cite{vshah_ATel}. Of the 22 bursts, twelve have only intensity data, while ten have both intensity and baseband data. Six baseband bursts were also recorded at KKO. The timeline of the burst detections are shown in Figure~\ref{fig:exposure} and the burst properties are listed in Table~\ref{tab:burst_prop} in Appendix \ref{sec:Appendix_burst_properties}, which also provides details about how the burst properties were estimated. Figure~\ref{fig:wfall} shows the dynamic spectrum of the ten bursts with baseband data, along with a frequency channel integrated time series, and a time integrated spectrum. These burst profiles were obtained by beamforming the baseband data to the VLBI position of \frb{} (see Table \ref{tab:VLBI_pos}). It is evident that most bursts from this repeater are narrowband, having a fractional bandwidth of $20-50$\% within the CHIME observing band. We note that following the burst activity in July, we did not detect any bursts from this repeater until another cluster of activity in October 2024. However, we limit the analysis in this paper to the bursts detected until 2024 July 31. 

Following the announcement of the discovery of \frb{} during its heightened activity in June 2024 \citep{vshah_ATel}, the Northern Cross FRB collaboration reported the discovery of a single burst detected during a follow-up campaign \citep{Northern_Cross_ATel}. The burst was detected on 2024-07-02 at 21:24:24.540 UTC at 400\,MHz. CHIME/FRB also detected a burst on the same day at 08:36:06 UTC at 400\,MHz. \cite{Casey_Law_ATel} followed up the source with the Very Large Array in two, 2-hour segments on 2024-07-02 and 2024-07-05 and did not detect any bursts at the observing frequency of 1 -- 2\,GHz. During the heightened activity of \frb{} in October 2024, \cite{Omar_ATel} detected a single burst at 1.3 GHz using the Westerbork RT-1 telescope after 350 hours of observations.  

\begingroup
\setlength{\tabcolsep}{0.1pt}
\begin{figure*}
\begin{tabular}{cccc}
\includegraphics[width = 0.23\textwidth]{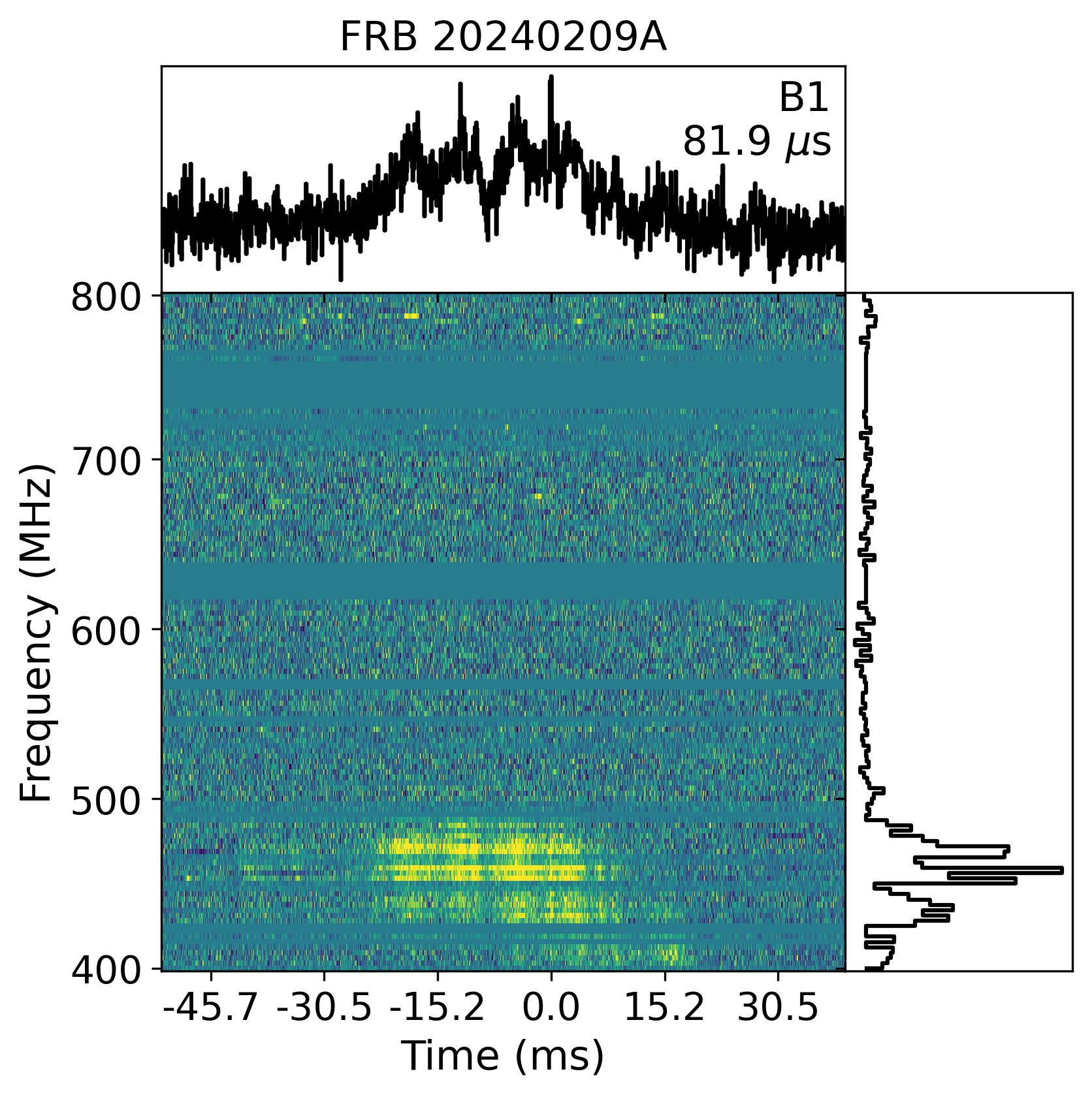} &
\includegraphics[width = 0.23\textwidth]{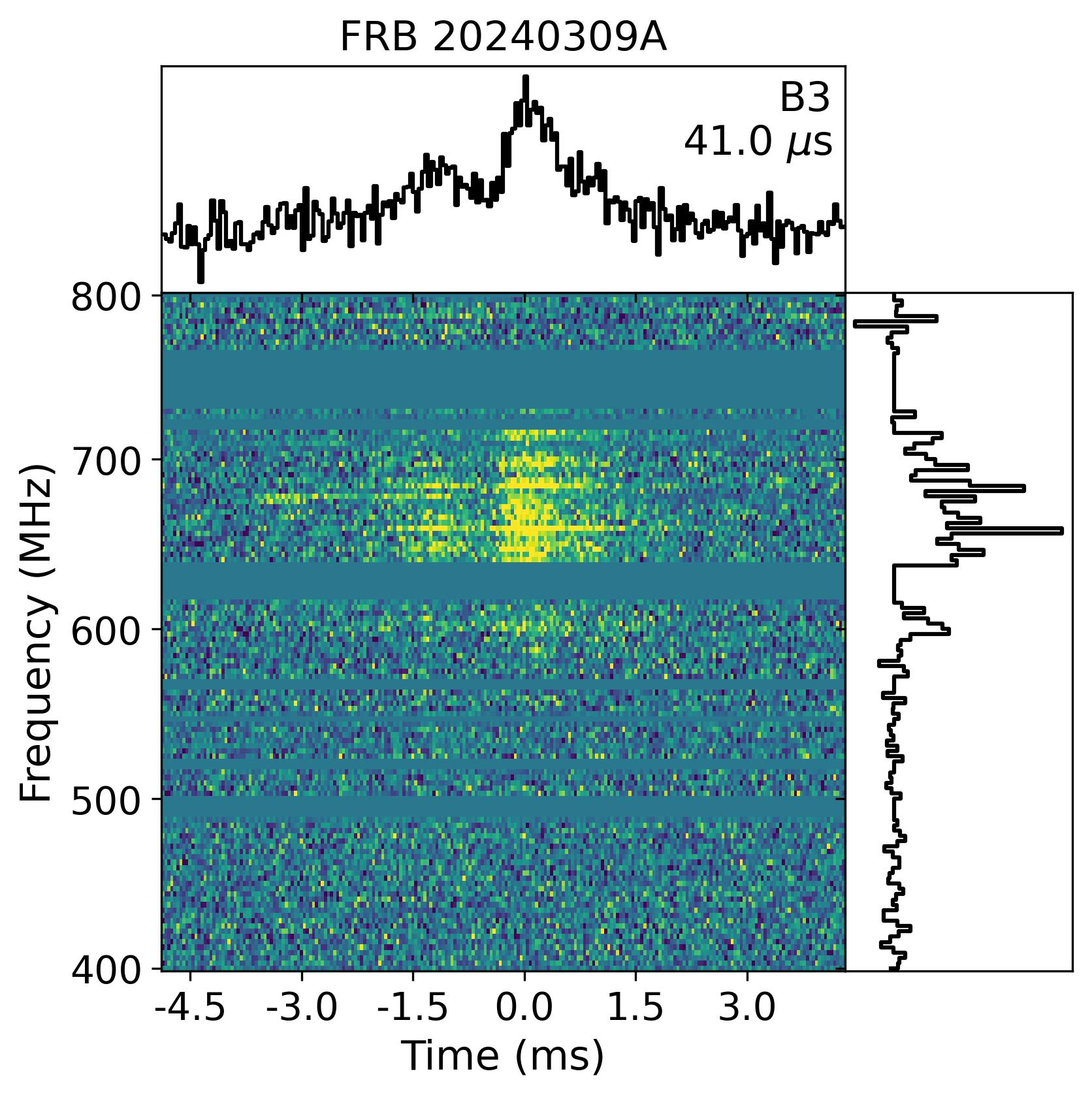} &
\includegraphics[width = 0.23\textwidth]{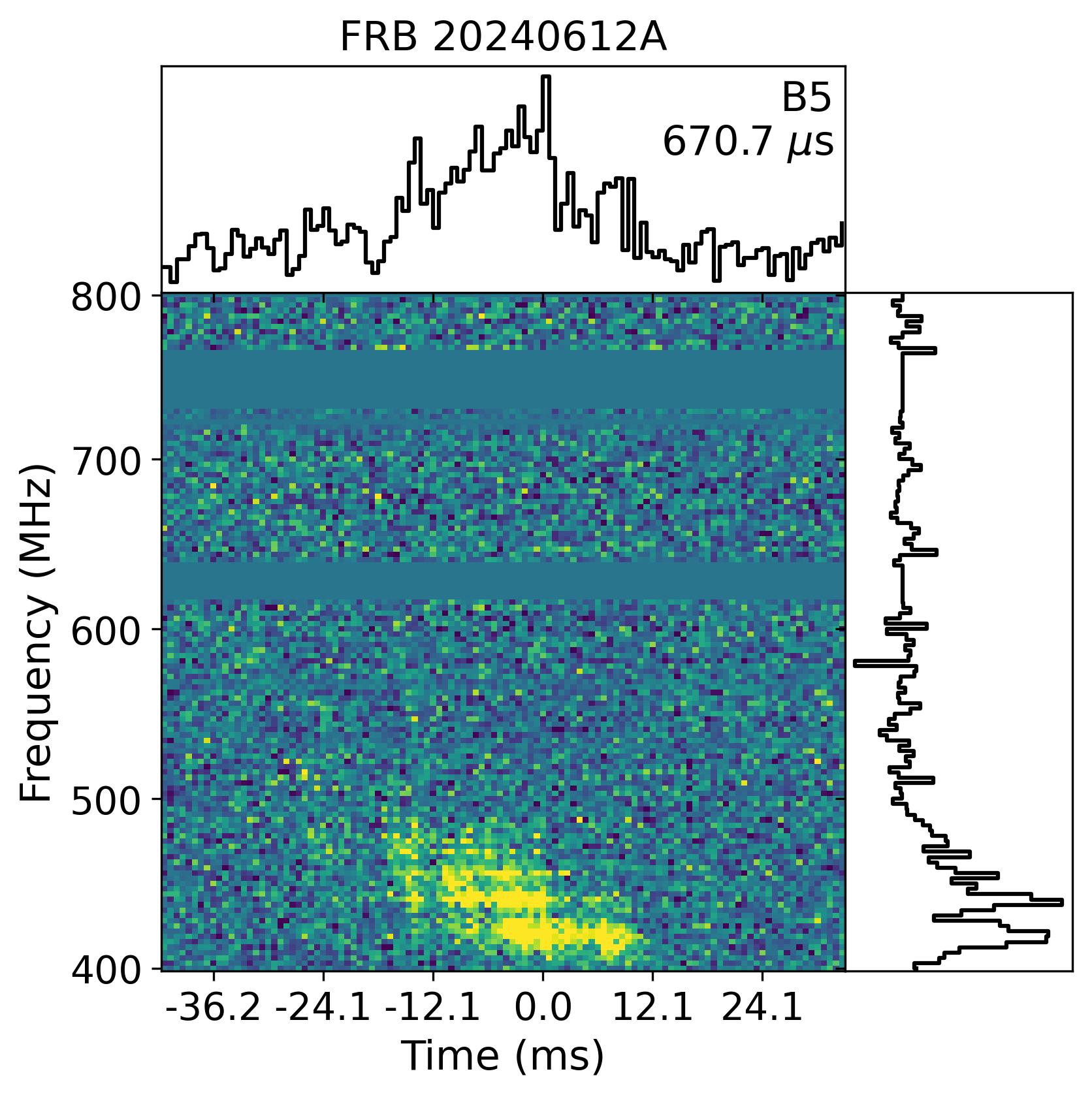} &
\includegraphics[width = 0.23\textwidth]{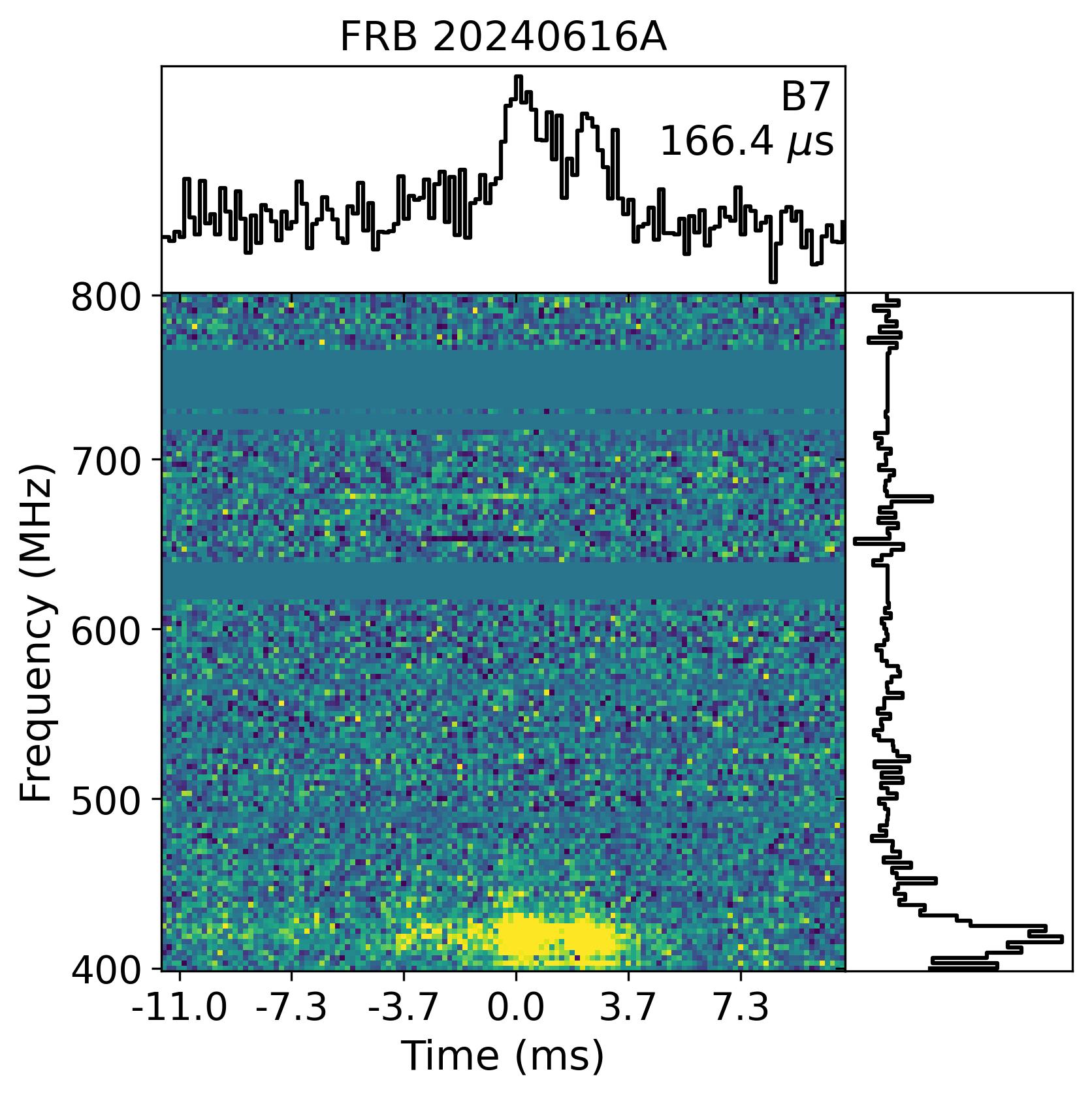} \\
\includegraphics[width = 0.23\textwidth]{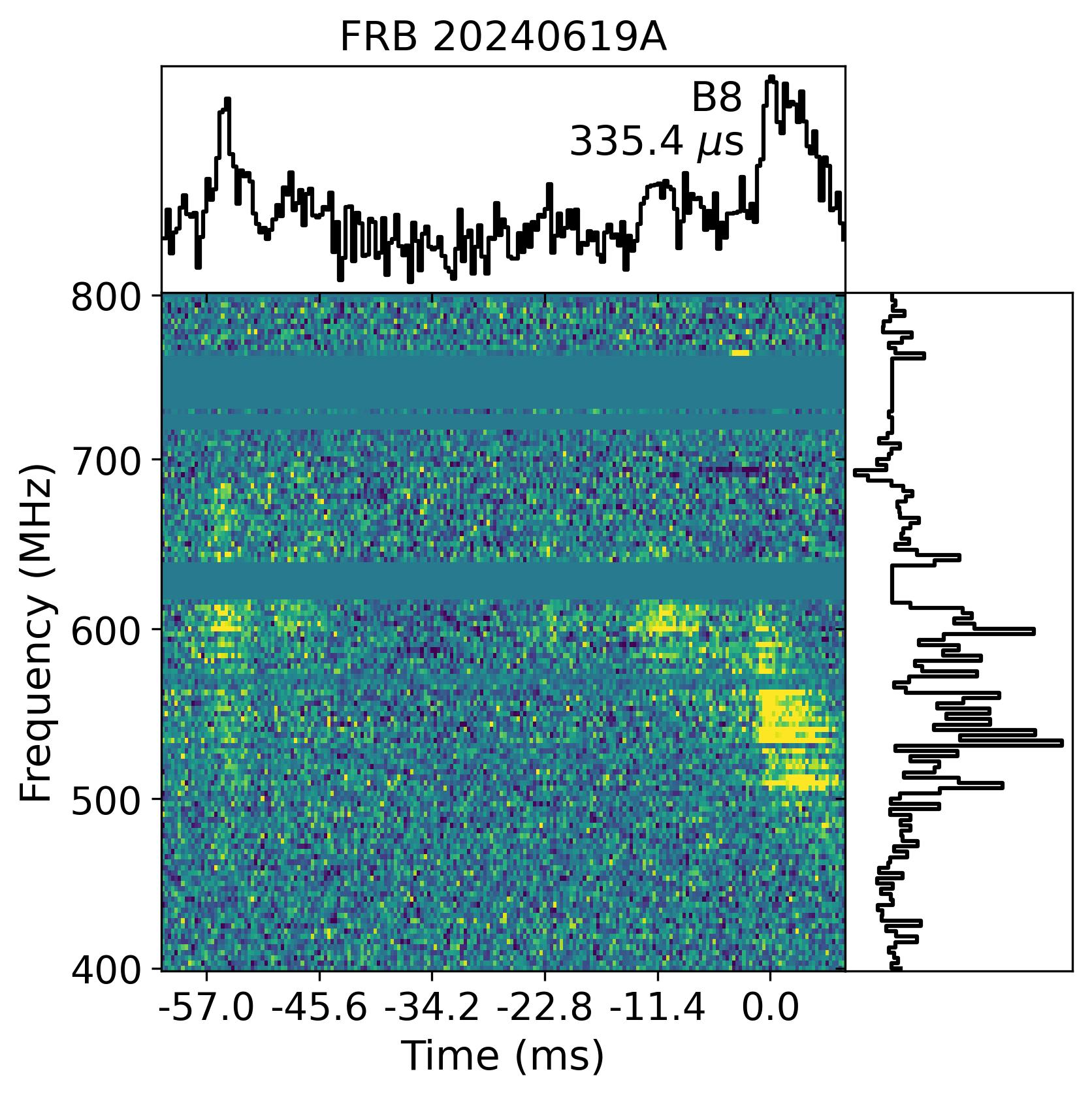} &
\includegraphics[width = 0.23\textwidth]{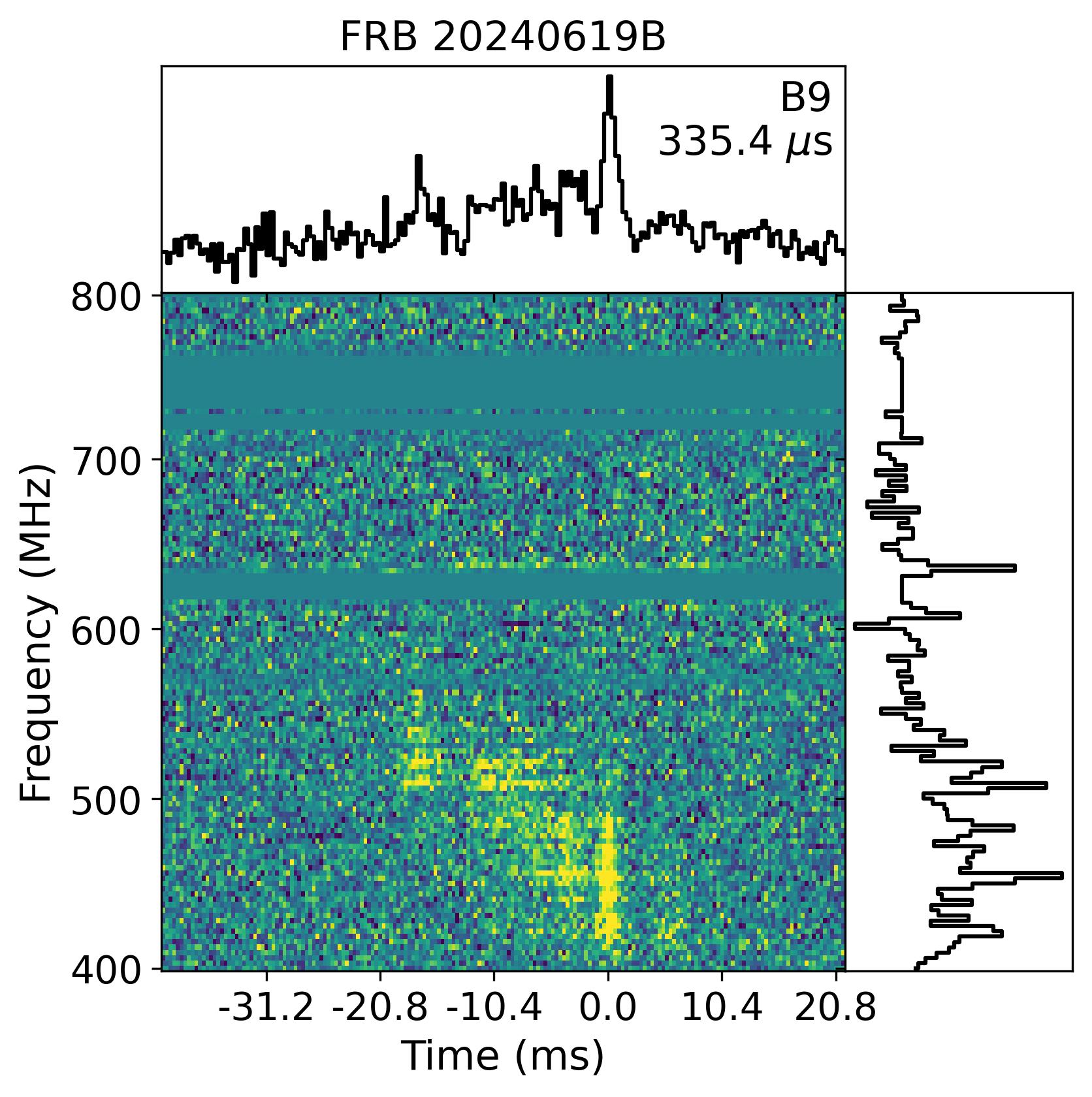} &
\includegraphics[width = 0.23\textwidth]{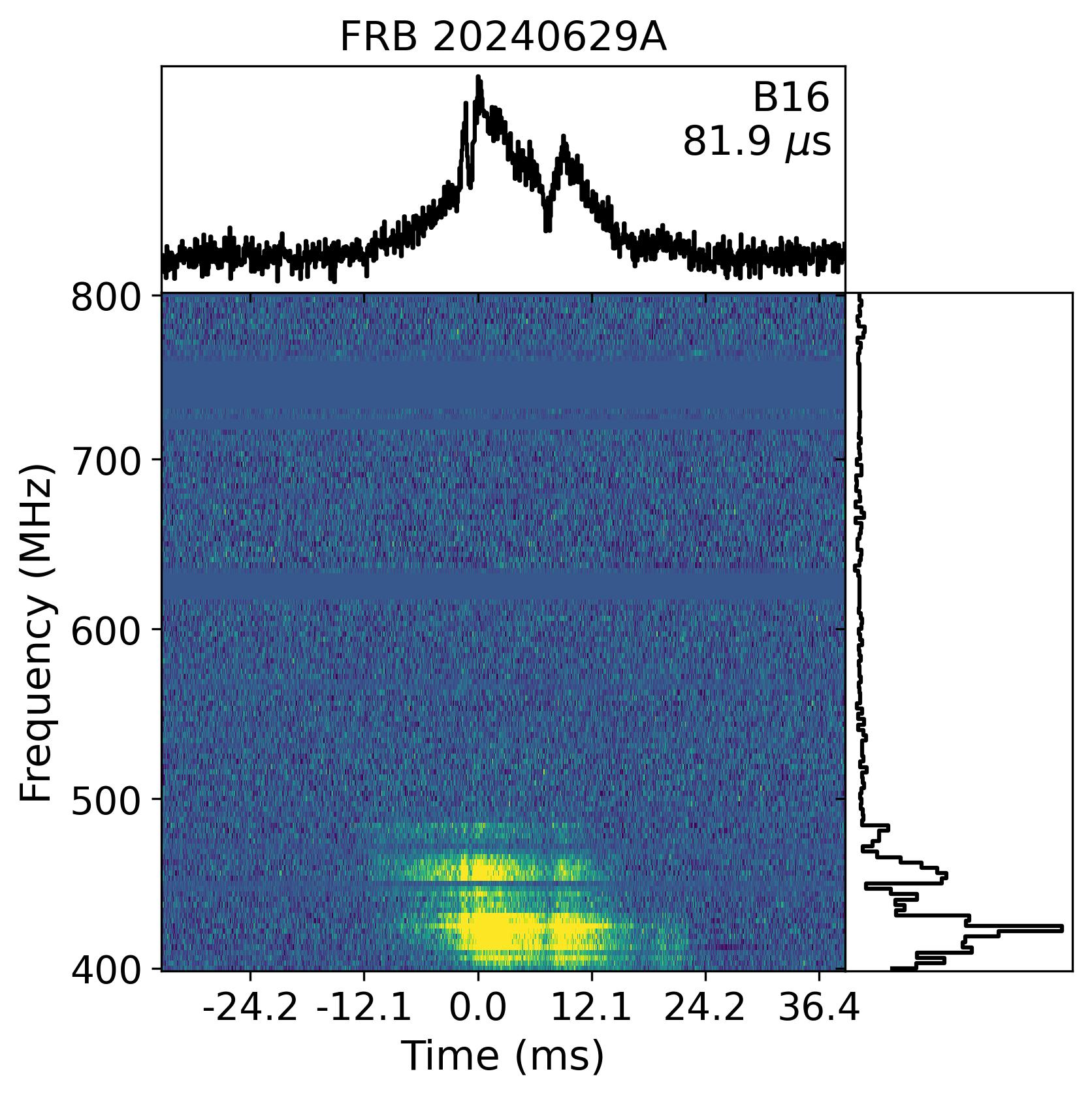} &
\includegraphics[width = 0.23\textwidth]{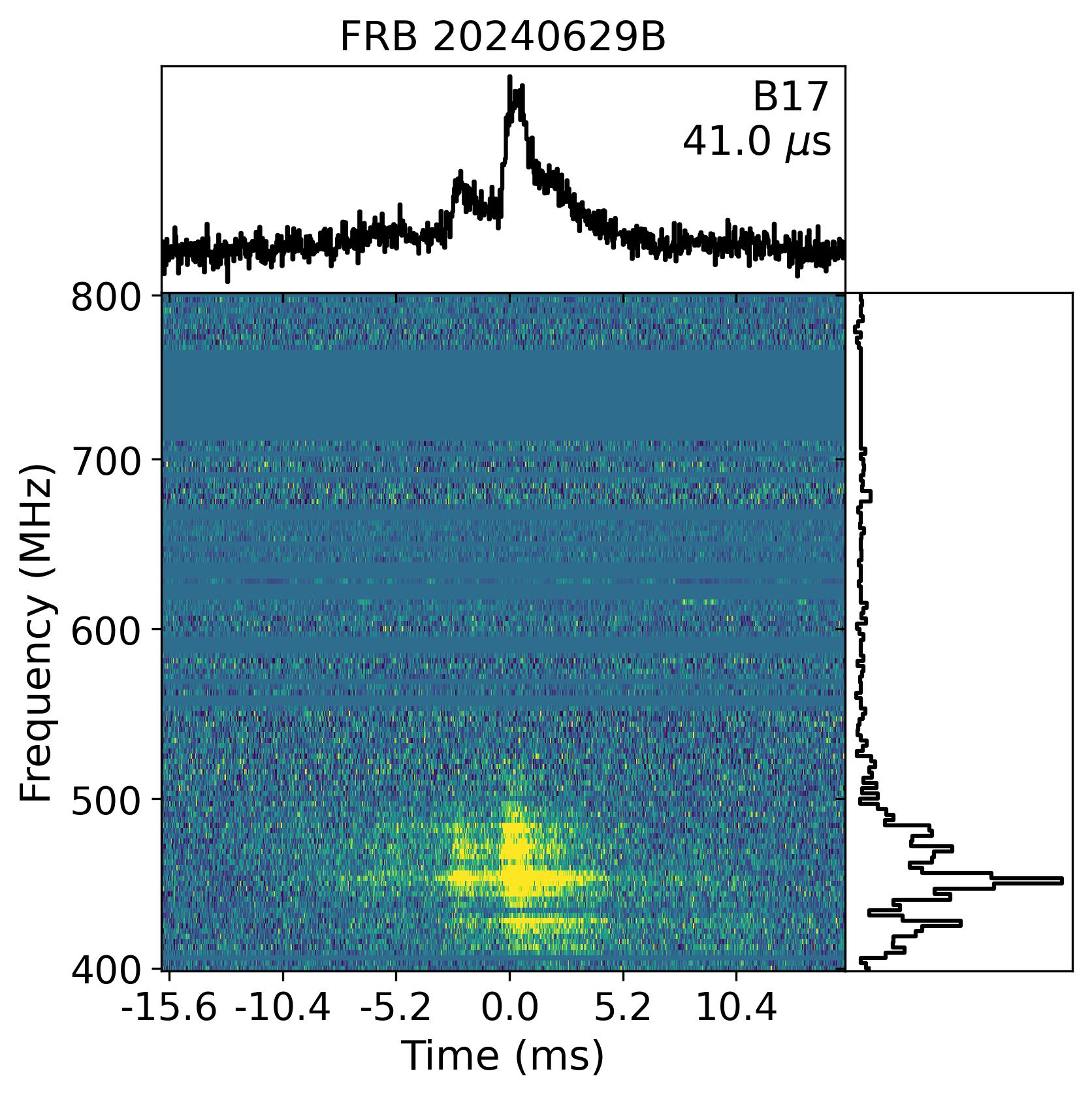} \\
\includegraphics[width = 0.23\textwidth]{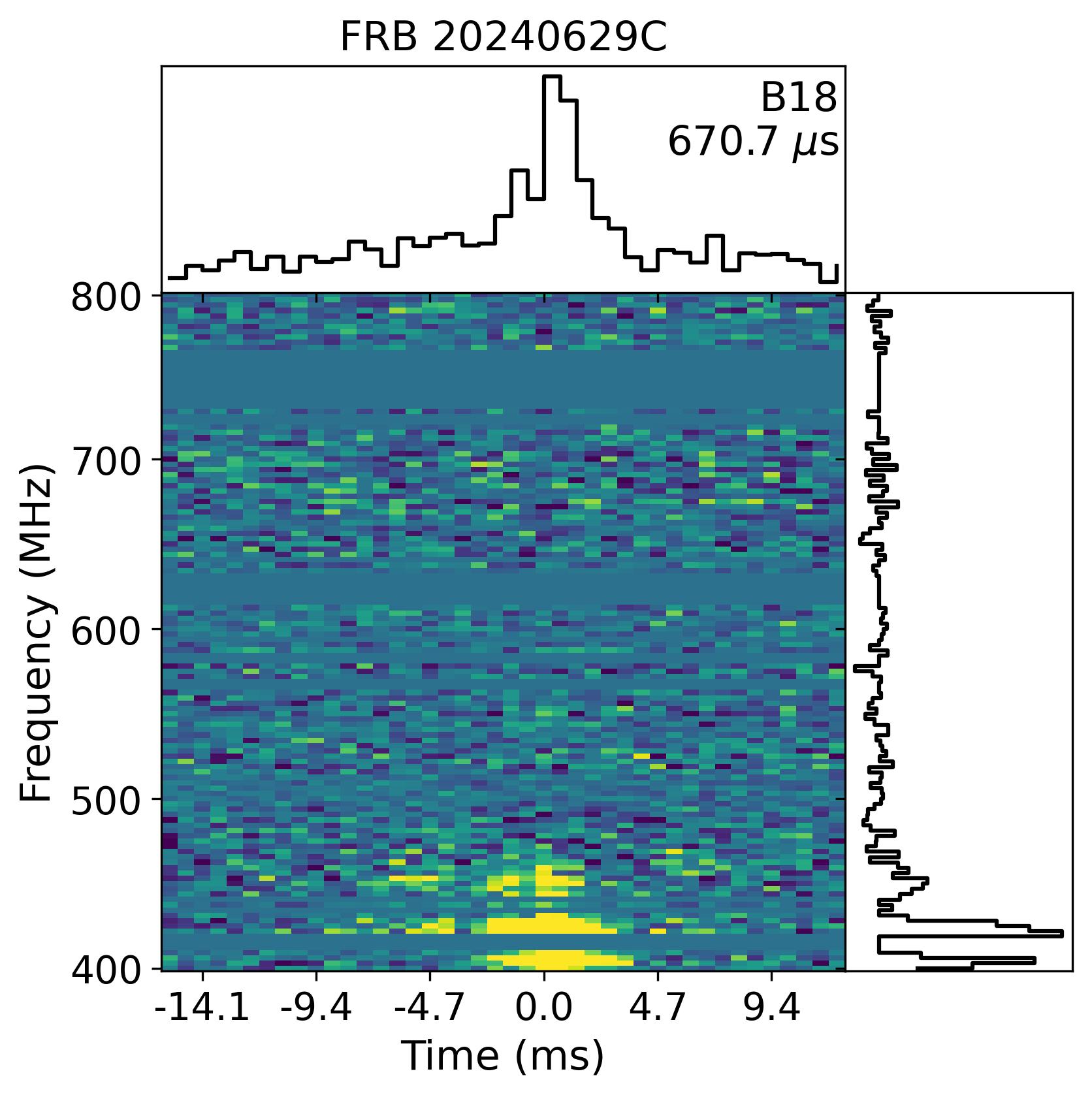} &
\includegraphics[width = 0.23\textwidth]{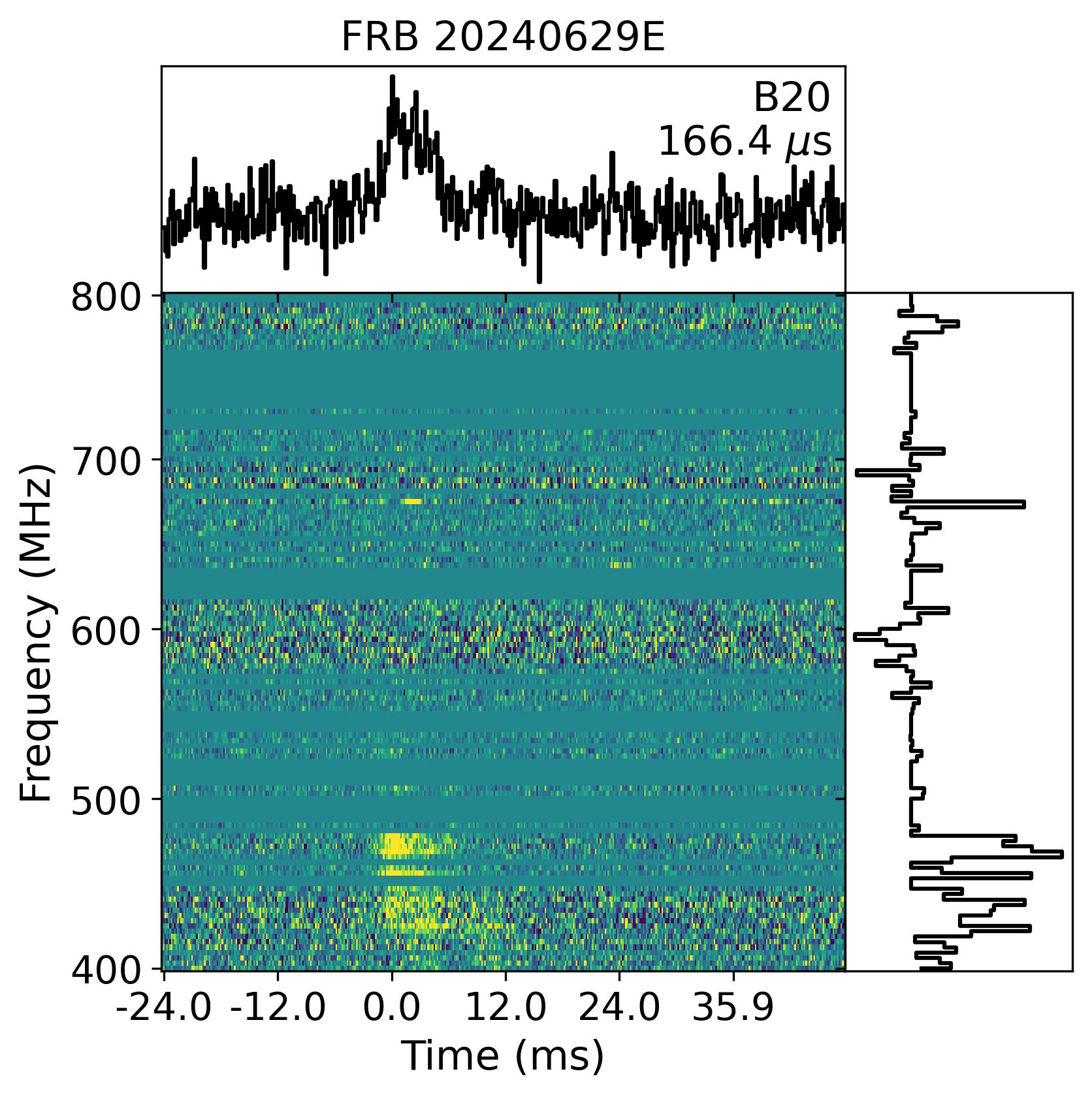} \\
\\
\end{tabular}
\caption{Dynamic spectra, frequency integrated profiles, and time integrated spectra of the ten repeat bursts from \frb{} that have baseband data saved at CHIME. The time resolution at which each burst is plotted along with the burst number (see Table \ref{tab:burst_prop}) are indicated in the top-right corner, and the TNS name of the burst is provided in the heading of each plot. }
\label{fig:wfall}
\end{figure*}
\endgroup

\subsection{Burst Rate Estimate} 
The total exposure to the source was estimated using the functionality described by \cite{CHIME_catalog_1}. Various metrics throughout different data processing stages were recorded to characterize the total uptime and variation in sensitivity of the entire CHIME/FRB system. This information was combined with the beam model\footnote{\url{https://chime-frb-open-data.github.io/beam-model/}} in order to account for times when a source transits through the full width at half maximum (FWHM) of our formed beams at 600 MHz. 

The exposure was queried using the best-known position of \frb{} \citep{vshah_ATel}. Due to the high declination of this source, it is circumpolar in CHIME's field-of-view, i.e., it has two transits separated into upper and lower transits \citep{CHIME_catalog_1}. It is important to note that the two transits have different sensitivities for the same source. CHIME's exposure to this source from 2018 August 28 to 2024 July 31 was $\sim$362 hours in the upper transit and $\sim$1123 hours in the lower transit, amounting to a total of $\sim$1485 hours.

The first burst was detected in the lower transit of the source giving a rate of $<$0.001\,hr$^{-1}$ at a 95\% fluence completeness threshold of 0.9\,Jy\,ms. The completeness threshold was estimated using the methodology described by \citet{CHIME_catalog_1} and \citet{Alex_R1}. Following the first detection, CHIME/FRB detected 13 repeat bursts in the upper transit and 8 repeat bursts in the lower transit in $\sim$34 hours and $\sim$90 hours of exposure, respectively, giving a burst rate\footnote{The burst rates are upper-limits because they also include the bursts detected outside the FWHM of our formed beams at 600 MHz} of $<$0.4\,hr$^{-1}$ at a 1.5\,Jy\,ms threshold and $<$0.1\,hr$^{-1}$ at a 0.9\,Jy\,ms threshold in the two transits. During its peak activity in CHIME on 2024 June 29th, 4 bursts were detected over a single, $\sim$13 min upper transit. This amounts to a rate of $<$20\,hr$^{-1}$, potentially implying an increase as high as $10^{4}$ times its initial activity in the upper transit. This increase in detection rate signifies that the source entered a high activity period fairly recently, and more broadly that some FRBs may rapidly transition from years of quiescence to hyperactivity on timescales of weeks to months as shown in Figure \ref{fig:exposure}. Such burst activity, from quiescence to hyperactivity, has also been seen for sources such as FRB~20201124A \citep{Adam_R67} and FRB~20220912A \citep{McKinven2022ATel_R117}. Additionally, as shown in Figure \ref{fig:exposure}, CHIME had limited exposure in the direction of this repeater following its peak burst activity because the FRB search engine gets shut down when the temperature in the computing container becomes too high during summer months. Thus, there is a possibility that additional bursts from this source were missed during its transit in July.

\begin{figure*}[t]
\centering
\includegraphics[width=0.7\textwidth]{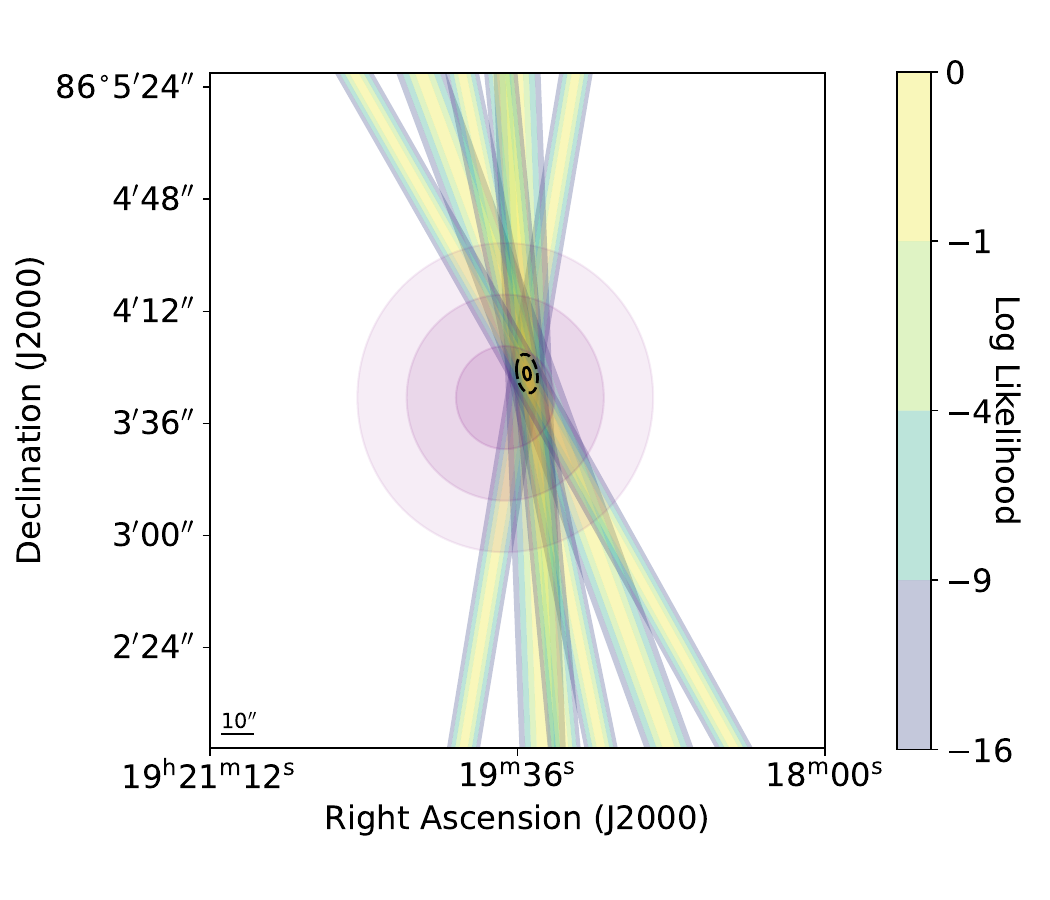}
\caption{The purple shaded ellipses show the 1-$\sigma$, 2-$\sigma$, and 3-$\sigma$ CHIME-only baseband localization region for \frb. The contour stripes show the CHIME--KKO VLBI localization regions for six repeat bursts, each constrained along the baseline vector. The black ellipses show the 1-$\sigma$ and 3-$\sigma$ combined VLBI localization region after accounting for correlated systematic errors.\label{fig:all_locs}.}
\end{figure*}

\section{Localization \label{sec:localization}}
Baseband data for six \frb{} bursts were saved at the KKO site, allowing six separate CHIME--KKO VLBI\footnote{The 66-km long CHIME--KKO baseline is not a `very-long' baseline per se. However, VLBI can often mean a non-connected interferometer where the voltage data need to be digitally shipped and correlated retrospectively. We use the term `VLBI' in that context in this paper.} localizations for this repeater. For each of these bursts, full-array baseband data were beamformed towards the best-fit FRB position found from the baseband localization pipeline, RA (J2000) = 19:19:39.84, Dec (J2000) = +86:03:44.28 \citep{baseband_loc, vshah_ATel}, as well as the position of the source J0117+8928 from the Very Long Baseline Array (VLBA) Radio Fundamental Catalog\footnote{ \url{https://astrogeo.org/sol/rfc/rfc_2024a/}}, which was used as the phase calibrator. This beamforming was done for both the CHIME and KKO stations. J0117+8928 was the best choice of calibrator, as it is unresolved on the CHIME--KKO baseline, and detected with a cross-correlation S/N $>$ 50. It also has a small angular separation of $\sim4^{\circ}$ from the FRB position, and is always within the primary beam of the CHIME and KKO cylinders because of its very high declination of 89.47$^{\circ}$. For each of the six target bursts, CHIME and KKO data were coherently de-dispersed using a DM of 176.518\,pc\,cm$^{-3}$, cross-correlated using the \texttt{PyFX} VLBI correlator \citep{PyFX}, and then phase referenced to the cross-correlated visibilities of J0117+8928. The target visibilities were integrated for the duration of the burst while the calibrator visibilities were integrated for the entire $\sim$100\,ms duration of the baseband dumps. The calibrated visibilities were obtained for both the E-W and N-S polarizations. Since the phase calibrator data were recorded simultaneously to the burst data, no clock calibration was necessary \citep{leung2021synoptic}. Though CHIME and KKO view the source through different sightlines through the ionosphere, this is not a dominant source of error for the short CHIME--KKO baseline at the observing frequencies we use at 400 -- 800 MHz \citep{KKO_overview}. Moreover, the small angular separation between the target and the calibrator ensures that the telescopes are looking through the same isoplanatic patch, such that there are no appreciable differential ionospheric phase delays between the target and the calibrator sightlines \citep{KKO_overview}. Only the visibility data for frequency channels that had signal were used for localization, which amounted to $60-100$\,MHz of bandwidth used for the narrowband bursts from this repeater.

A grid of RA and Dec centered on the baseband localization was searched to determine the CHIME--KKO VLBI localization position using a `delay mapping' technique \citep[see Equations~12, 13 and 19 of][]{KKO_overview}. The differential geometric delay between CHIME and KKO with respect to the initial pointing was estimated for each point in the grid and compared to the differential geometric delay obtained from the calibrated visibilities, which gave a localization likelihood \citep[see Equation~A1 of][]{PyFX}. The uncertainty on the localization was determined by the uncertainty on the delay, which is typically $\sim$1\,ns for the CHIME--KKO baseline \citep{KKO_overview}. However, since the bursts from this repeater are narrowband, a more conservative delay uncertainty of 2\,ns (assumed to be Gaussian distributed) was used to account for unknown systematic errors and residual ionospheric delays \citep[see Figure~13 of][]{KKO_overview}. If the delays obtained from the E-W and N-S calibrated visibilites differed by more than 1\,ns, the polarization with the higher cross-correlation S/N was used for localization. Localization likelihoods for one or more polarizations for each of the six bursts were multiplied to obtain the combined localization. This localization region was inflated by convolving it with a 2-D Gaussian having $\sigma = 1^{\prime\prime}$ to account for correlated systematic errors (see Appendix~\ref{Appendix_B} for more details), giving a final localization ellipse of dimensions $\sim 1^{\prime\prime}\times2^{\prime\prime}$. 

In Figure~\ref{fig:all_locs}, the purple ellipses show the 1-$\sigma$, 2-$\sigma$, and 3-$\sigma$ CHIME-only baseband localization regions for \frb{} \citep{vshah_ATel}. The contour stripes show the six CHIME--KKO localizations, with the solid and dashed black ellipses delimiting the 1-$\sigma$ and 3-$\sigma$ combined localization regions, respectively. The high declination of this repeater enabled its detection with a rotating range of baseline vectors, permitting the combined VLBI localization region to be constrained along several axes. The parameters for the CHIME--KKO 1-$\sigma$ localization ellipse are listed in Table \ref{tab:VLBI_pos}.

\begin{deluxetable*}{ c c c c c }
\tabletypesize{\small}
\tablehead{
\colhead{Right Ascension} & \colhead{Declination} & \colhead{Semi-minor axis} & \colhead{Semi-major axis} & \colhead{Position Angle} }
 \startdata
{$19^{\mathrm{h}}19^{\mathrm{m}}33^{\mathrm{s}}$} & {$+86^{\circ}03^{\prime}52^{\prime\prime}$} & {$1.08^{\prime\prime}$} & {$2.12^{\prime\prime}$} & {$9.54^{\circ}$}
\enddata
\caption{Parameters for the 1-$\sigma$ CHIME--KKO VLBI localization ellipse for \frb, with the center of the ellipse defined in the ICRS J2000 reference frame. \label{tab:VLBI_pos}}
\end{deluxetable*}

\section{Host Galaxy \label{sec:hostgalaxy}}
\subsection{Deep imaging and host galaxy association}
We imaged the field of \frb{} with the Gemini Multi-Object Spectrograph (GMOS) mounted on the 8-meter Gemini-North telescope on UT 2024 August 8 (PI: T.~Eftekhari), for a total exposure of 1~hr in $r$-band. We used the \texttt{POTPyRI pipeline}\footnote{\url{https://github.com/CIERA-Transients/POTPyRI}} to apply bias and flat-field corrections, and co-add the images. We performed astrometric calibration using sources in common with the {\it Gaia DR3} catalog with an astrometric tie RMS of $0.06\arcsec$. We performed aperture photometry of the host galaxy within a $14\arcsec$ radius aperture and calculated a magnitude $r = 16.79 \pm 0.02$~mag (AB). 

\begin{figure}
\centering
\includegraphics[width=\columnwidth]{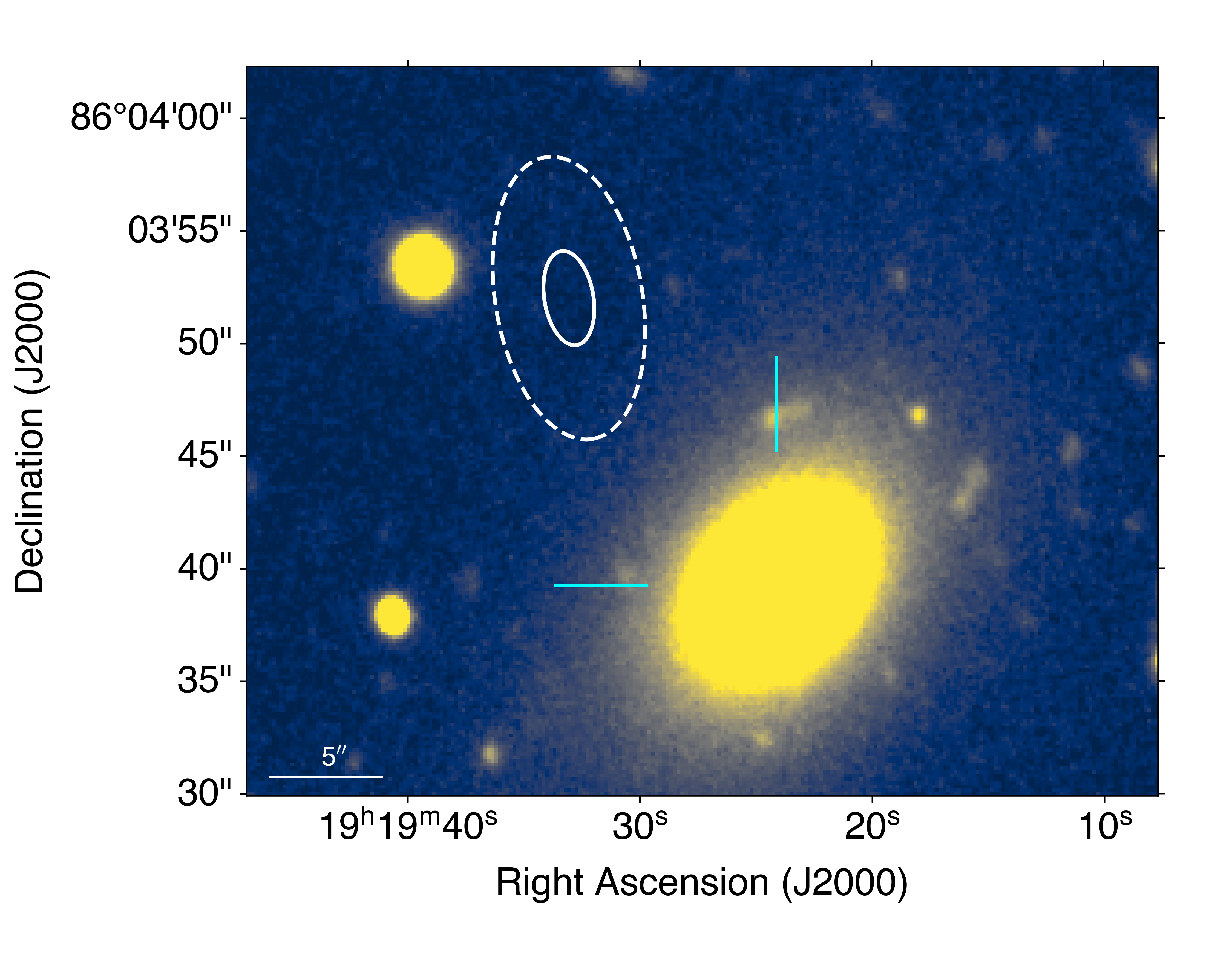}
\caption{Gemini $r$-band image showing the host galaxy of \frb{} (cyan crosshairs) and the 1-$\sigma$ (white solid) and 3-$\sigma$ (white dashed) localization ellipses. The source to the East of the localization ellipse is a star. The 3-$\sigma$ limiting magnitude is $r \gtrsim 25.9$~mag (corrected for Galactic extinction). \label{fig:deep_gemini}}
\end{figure}

To determine the most probable host galaxy, we utilized the Probabilistic Association of Transients to their Hosts method (PATH; \citealt{Aggarwal2021}), a Bayesian framework for associating transients with their hosts. 
We adopted the default ``inverse'' prior that accounts for the higher incidence of faint galaxies on the sky (i.e.,\ a Jeffreys prior for apparent magnitude) and an underlying exponential distribution scaled by one-half the half-light radius as the prior for the offset distribution \citep{craft_ics_paper}. For the prior probability that the host is unseen, we assumed a conservative value of $P(U) = 0.15$ given the 3-$\sigma$ limiting magnitude of the Gemini/GMOS image of $r \approx$ 25.9 mag (AB, corrected for a Galactic extinction of $A_r = 0.266$~mag). To identify galaxy candidates, we utilized \texttt{Source Extractor} on a $40\arcsec \times 40\arcsec$ region centered on the FRB position. We performed aperture photometry on each of the identified sources, and fed as input to PATH the extinction-corrected magnitudes, source positions, and estimates for the angular sizes. The host of \frb{} was robustly identified with a posterior probability $P(O|x) = 0.99$ (Figure~\ref{fig:deep_gemini}).

Using the spectrum of the host galaxy, we classify it as a quiescent galaxy at $z = 0.1384 \pm 0.0004$ \citep{Eftekhari_R155}, adding to the small sample of FRB hosts in quiescent environments \citep{Bannister2019,DSA_quiescent,Law2024,Sharma2024}. The properties of \frb{} are especially notable given no FRB associated with a quiescent galaxy has yet been observed to repeat. \frb{} is also the first FRB localized to a galaxy with unambiguous elliptical morphology \citep{Eftekhari_R155}, with only one other non-repeating FRB having a candidate elliptical host \citep{Sharma2024}. The global host galaxy properties, especially its extremely old stellar population ($\sim 11$  Gyr; \citealt{Eftekhari_R155}), coupled with the large host galaxy offset evident in Figure \ref{fig:deep_gemini}, support a delayed formation channel for the progenitor of \frb. We discuss these implications in Section~\ref{sec:discussion}.

In order to verify the consistency of the host galaxy redshift with the observed DM of \frb{}, we use the joint probability distribution of redshift and extragalactic DM for CHIME repeaters, developed by \citet{james2023_chime_zDM}, to estimate a probabilistic maximum redshift. We use the NE2001 estimate of 55.5\,pc\,cm$^{-3}$ for the MW disk DM contribution \citep{NE2001, Ocker_NE2001}, which is consistent with the YMW16 estimate of 52.2\,pc\,cm$^{-3}$ \citep{YMW16}. We assume both the MW halo DM contribution (DM$_{\mathrm{MW, halo}}$) and host galaxy DM contribution DM$_{\mathrm{host}}$ are 0\,pc\,cm$^{-3}$ in order to obtain a more conservative $z_\mathrm{max}$. Given that the extragalactic DM is the total DM minus the NE2001 contribution, the resulting 95\% upper limit on redshift is $z_\mathrm{max} = 0.19$. Thus, the redshift of the putative host galaxy is consistent with the probabilistic maximum redshift given by the DM of \frb{} but suggests a small contribution to the DM from the local environment and host galaxy of the FRB. 

\begin{figure}[ht!]
\centering
\includegraphics[width=0.5\textwidth,clip=]{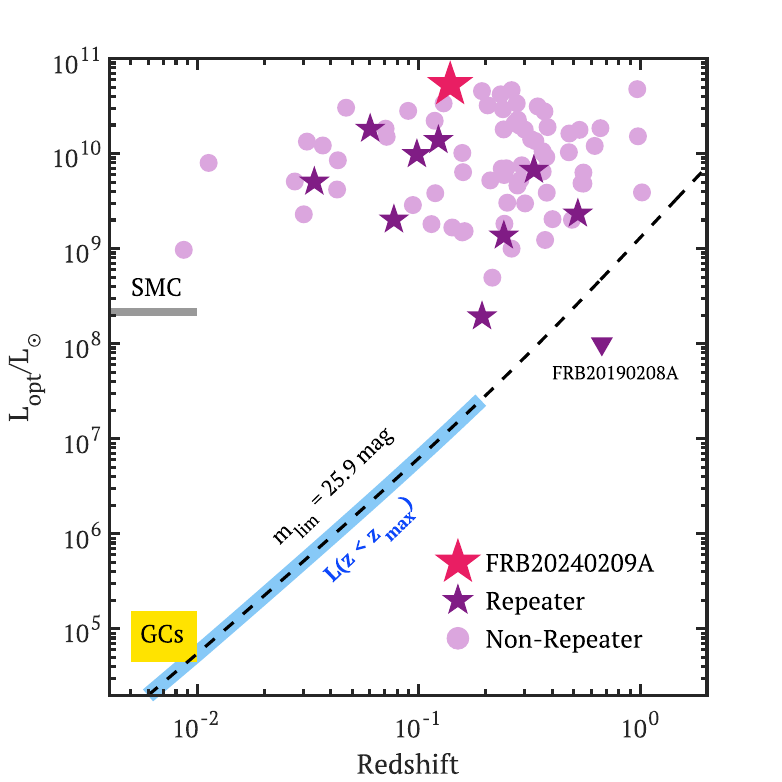}
\caption{Optical luminosity versus redshift for the host galaxy of \frb{} (red star) and other FRB hosts in the literature \citep{Gordon_2023,Bhardwaj_2024,Ibik_2024,Lee-Waddell_2023,Ravi_2019Nature,Law2024,Shannon2024,Sharma2024,Connor2024}; stars denote known repeaters while circles denote apparent non-repeaters. The dashed line corresponds to the Gemini luminosity limit of $r \approx 25.9$~mag (3-$\sigma$), for which parameter space above the line is ruled out for a source coincident with the KKO localization. The range of luminosity limits on a coincident source, set by the upper limit on the redshift inferred from the DM ($z_{\rm max} = 0.19$), is highlighted in blue. We place a range of $\approx 3 \times 10^{5}\,L_{\odot}$ to $\approx 2.5 \times 10^{7}\,L_{\odot}$ for a coincident source. This is well below the luminosity of the Small Magellanic Cloud (gray horizontal line), and below the limit of the inferred luminosity for the host of FRB\,20190208A \citep{Dante_dwarfhost}; purple triangle). The lower bound reaches the luminosities of globular clusters in elliptical galaxies (yellow box; e.g., \citealt{Strader_2006}). 
}\label{fig:lumz}
\end{figure}

To place the host of \frb{} into context with the FRB host population, in Figure~\ref{fig:lumz} we show the optical $r$-band luminosities\footnote{In a few cases where $r$-band is not available, we use $i$-band.} of FRB host galaxies as a function of redshift; all magnitudes are corrected for Galactic extinction \citep{Gordon_2023,Bhardwaj_2024,Ibik_2024,Lee-Waddell_2023,Ravi_2019Nature,Law2024,Shannon2024,Sharma2024,Connor2024}. We also include the inferred limit on the luminosity for the low-lumionsity host of FRB\,20190208A \citep{Dante_dwarfhost}. With $L \approx 5.3 \times 10^{10}\,L_{\odot}$, the putative host galaxy of \frb{} is the most luminous FRB host galaxy to date, three times more than the most luminous host of a known repeating FRB (although only $\approx 10\%$ more luminous than that of a non-repeating FRB). Here we entertain the possibility that there exists an undetected source within the combined KKO localization. Our deepest Gemini imaging reaches a 3-$\sigma$ limit of $r \approx 25.9$~mag derived using the image FWHM of $0.6\arcsec$, which we translate to the required combination of luminosity and redshift to exist below our detection threshold (Figure~\ref{fig:lumz}). Given the redshift constraints from DM and the location of the FRB, one possibility is that of a satellite of the putative host at the same redshift ($z=0.1384$), for which the Gemini limit requires that $L \lesssim 1.2 \times 10^{7}\,L_{\odot}$. Alternatively, there may exist a galaxy at a redshift different from that of the putative host. For $z<z_{\rm max}$ where $z_{\rm max}=0.19$ (95\% confidence) the Gemini limit requires $L \lesssim 2.5 \times 10^{7}\,L_{\odot}$. Even for the most optimistic case of $z=z_{\rm max}$, our limits imply a luminosity $\gtrsim 10$ times less than that of any other known FRB host galaxy with a redshift \citep{Gordon_2023,Bhandari_2023}. We note that the dwarf host of FRB\,20190208A does not have a redshift, but has a range of luminosities of $\approx 10^{7}-10^{8}\,L_{\odot}$ inferred from the DM \citep{Dante_dwarfhost}, potentially comparable to the limit for \frb{}. Thus, if \frb{} originated from an undetected galaxy, it would need to be extreme in terms of its low luminosity compared to the rest of the FRB population. We discuss this in the context of possible progenitors in Section~\ref{sec:discussion}.

\section{Discussion \label{sec:discussion}}
The burst properties, activities, and counterparts of FRBs give clues about their emission mechanisms, progenitor types, and local environments. We compare these attributes of \frb{} to those of other repeating FRBs. Moreover, the host galaxy properties and location of the FRBs within their hosts help identify the most viable FRB formation channels possible for specific environments. We therefore discuss the implications for the progenitor of \frb{} based on its offset from the putative host. 

\subsection{Comparison with other repeaters \label{subsec:burst_prop_comp}} 

\subsubsection{Burst properties}
The wide temporal widths and narrowband nature of \frb{} bursts are consistent with what is observed in the broader repeater population \citep{Ziggy_repeater_morphology}. Moreover, bursts B1, B5, B8, B9, and B16 clearly show a downward-drifting morphology in the frequency-time space. Among repeaters, FRB~20121102A is known to emit isolated bursts of microsecond duration \citep{Snelders_R1_microshots}, while FRB~20200120E shows substructure down to a timescale of $\sim$60\,ns in sub-ms duration bursts \citep{DSN_M81R, Kenzie_M81R}. If \frb{} emitted such narrow bursts, they would have to be brighter than the wider bursts in order to be detected by the CHIME/FRB backend, which detects with a time resolution of $\sim$1\,ms. Moreover, the bursts from FRB~20200120E are 2-3 orders of magnitude less energetic than those from other repeating FRBs, and thus easily detectable for FRB~20200120E only due to the proximity of this source located 3.6\,Mpc away. Although \frb{} does not show such isolated narrow bursts, it does show evidence of structures of varying timescales within the broader burst envelope, especially in burst B1. Such structures are also seen in bursts from other repeaters \citep{Kenzie_microshots, Dante_R117_microshots}. Thus, the morphological features of \frb{} are consistent with those of other repeaters and suggest that \frb{} potentially has the same emission mechanism and progenitor type as other repeating FRBs. However, this does not preclude the progenitor of \frb{} from having a different formation channel compared to other repeaters. 

The brightest burst from \frb{} in our sample is the first burst detected from this source, which has a structure-maximizing DM of 176.49\,pc\,cm$^{-3}$. The DMs of the repeat bursts listed in Table \ref{tab:burst_prop} are largely consistent. We do not consider any apparent DM variation to be intrinsic to the source as it is non-trivial to disentangle DM variation from downward drifting morphology, especially at the coarser time resolution for some bursts. Moreover, the DM variation of $>$1\,pc\,cm$^{-3}$ for FRB~20121102A occurs over a period of years \citep{R1_DM_variation}. Thus, significant DM variation is not expected in the \frb{} bursts listed here, which span a period of a few months. 

\subsubsection{Burst activity}

The daily exposure of CHIME towards the position of \frb{} places strong constraints on the burst rate before the source turned on, and allows us to monitor the activity of the source once it is active. \frb{} was first detected in February 2024 and had a sudden period of heightened burst activity in June 2024, with the burst rate rising to $<$20\,hr$^{-1}$, almost $10^4$ times its initial upper-limit. A sudden increase in burst activity has also been seen in other repeaters: the first example is from the CHIME/FRB observation of a period of high burst activity from the repeater FRB~20201124A, with the peak burst rate reaching up to 92\,day$^{-1}$ and 201\,day$^{-1}$ with a 90\% fluence threshold of 19\,Jy\,ms during this period in the CHIME band \citep{Adam_R67}. For this source, FAST saw a peak burst rate of 542\,hr$^{-1}$ \citep{FAST_R67_II} in their 1--1.5\,GHz observing band with a 90\% fluence threshold of $\sim$0.03\,Jy\,ms. For FRB~20200120E, \citet{Kenzie_M81_burst_storm} observed a period of burst activity with the Effelsberg telescope at 1.4\,GHz, where the source emitted 53 bursts within 43 minutes above a fluence of 0.04\,Jy\,ms. \citet{FAST_R117} observed FRB~20220912A with the FAST telescope and observed an event rate up to 390~hr$^{-1}$ with a 90\% fluence threshold of $\sim$0.014\,Jy\,ms. Most recently, FAST also observed FRB~20240114A to have a burst rate up to $\sim$500\,hr$^{-1}$ above a fluence of 0.015\,Jy\,ms, an increase in burst activity by over an order of magnitude compared to observations conducted just over a week prior \citep{FAST_R147ATel}. The sudden increase in activity of \frb{} is thus consistent with other repeaters, although a direct comparison of burst rates across different frequency bands and fluence thresholds requires a more detailed analysis. 

As discussed in Section \ref{sec:observations}, apart from the CHIME detected bursts, there are only two other reported detections of \frb. The Northern Cross FRB collaboration detected a single burst at 400\,MHz in 20 hours of observation \citep{Northern_Cross_ATel}, while the Westerbork RT-1 telescope detected a single burst at 1.3\,GHz in 350 hours of observation \citep{Omar_ATel}. This is seemingly in contrast to other repeaters that have been highly active across a wide frequency band --- all the aforementioned FRB repeaters, which had follow-up burst activity monitoring campaigns at $>$1\,GHz, were discovered by CHIME/FRB at 400 -- 800\,MHz. Other repeating FRBs with detections across various frequency bands include FRB~20121102A \citep[from $\sim$600\,MHz to $\sim$8\,GHz, though seemingly preferentially at higher frequencies;][]{Alex_R1, R1_BreakthroughListen} and FRB~20180916B \citep[from $\sim$110\,MHz to $\sim$8\,GHz][]{Pleunis+2021_R3_LOFAR, Bethapudi+2023_R3}. A deeper analysis is required to study the burst activity of \frb{} across different frequency bands.

\subsubsection{Limits on a persistent radio counterpart}

Only FRB~20121102A and FRB~20190520B are confidently associated with compact persistent radio sources (PRSs) that are coincident with their positions \citep{121102_loc_VLA, R1_twin_PRS}. FRB~20201124A has a coincident radio source, which is interpreted as a candidate PRS \citep{bruni2024nebularoriginpersistentradio} or star formation \citep{Dong_20201124A_SF}. We performed an archival search for compact persistent radio emission at the position of \frb, utilizing CIRADA\footnote{https://cirada.ca/} cutouts to extract a $1~\mathrm{arcmin}^2$ region surrounding the VLBI position. The VLA Sky-Survey Quicklook (VLASS-QL; \citealt{2020PASP..132c5001L}) was the only radio survey with available observations of the field of \frb{}. We found no evidence for any PRS emission within the 3-$\sigma$ localization regions above the 5-$\sigma$ RMS threshold of $600~\mu$Jy. Assuming $z = 0.1384$, we place a 5-$\sigma$ upper limit on the spectral luminosity of any PRS counterpart of $L_{3.0~\mathrm{GHz}} \lesssim 3\times10^{29}~\mathrm{erg~s}^{-1}~\mathrm{Hz}^{-1}$. In comparison to those available in the literature, a PRS akin to those associated with FRB~20121102A and FRB~20190520B, which fall within the spectral luminosity ranges of $L_{1.7~\mathrm{GHz}}\approx 1-3\times10^{29}~\mathrm{erg~s}^{-1}~\mathrm{Hz}^{-1}$ (\citealt{121102_loc_EVN}; \citealt{bhandari2023constraintspersistentradiosource}), would be on the cusp of the 5-$\sigma$ detection threshold provided by VLASS-QL. On the contrary, a lower-luminosity PRS akin to the candidate PRS associated with FRB~20201124A with $L_\nu \sim 5\times10^{27}~\mathrm{erg~s}^{-1}~\mathrm{Hz}^{-1}$ \citep{bruni2024nebularoriginpersistentradio} would fall below the detection threshold. 

\cite{Casey_Law_ATel} searched for a PRS at the \frb{} position using VLA, and found an unresolved radio source which is at or near the nucleus of the putative host. Given the large offset of \frb{} from the putative host, this radio source is not associated with the FRB, and instead may indicate the presence of a radio-loud AGN in the host as discussed in \citet{Eftekhari_R155}. 
While a lower luminosity PRS counterpart cannot be ruled out at this stage, \frb{} currently joins the group of well-localized repeating FRBs with no observed compact radio counterparts \citep{Ibik_PRS}. 

\begin{figure*}[t]
\centering
\includegraphics[width=0.9\textwidth]{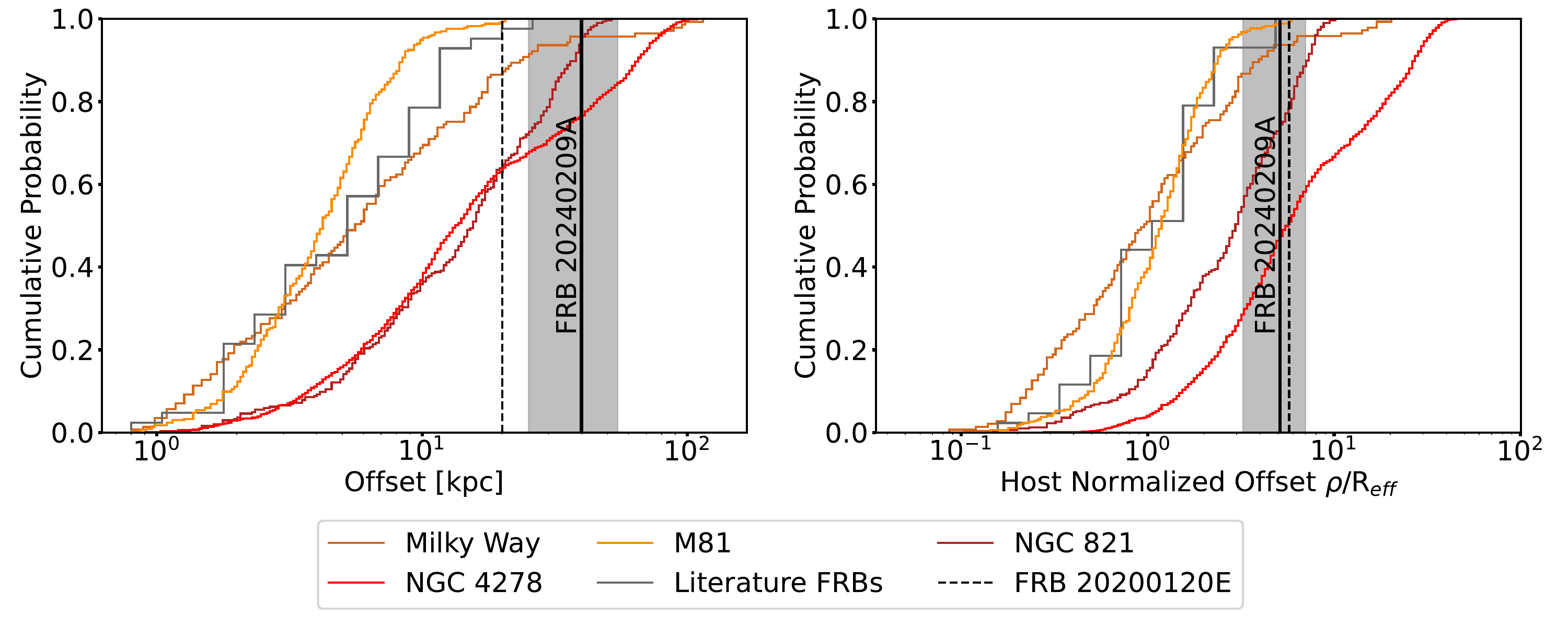}
\caption{The projected physical offsets (left panel) and host-normalized offsets (right panel) of forty-three FRBs which have robust host associations and offset information. Also shown are the projected physical and host-normalized offsets of globular clusters of the spiral galaxies Milky Way and M81, and of the elliptical galaxies NGC 821 and NGC 4278. The black dashed line shows the $\sim$20\,kpc projected offset of FRB 20200120E which is localized to a globular cluster of M81. The black solid line shows the $\sim$40\,kpc projected offset of \frb{} with the gray shaded region showing the 3-$\sigma$ uncertainty on the offset based on the localization region. Based on its offset, \frb{} is consistent with originating from a globular cluster of its massive elliptical host galaxy. }\label{fig:GC_offsets}
\end{figure*}

\subsection{Galactocentric offset and implications for the progenitor of \frb}

One of the most notable features about \frb{} is the offset from its host galaxy, which has implications for its progenitor. Thus, we use the combined CHIME--KKO localization to calculate the distribution of possible angular offsets between \frb{} and its host galaxy center, weighted by the localization probability map. Using $z=0.1384$ and Planck cosmology \citep{Planck_2018}, we calculate a projected physical offset of 40 $\pm$ 5\,kpc (68\% confidence). Since the host of \frb{} is particularly large in extent \citep{Eftekhari_R155}, it is informative to normalize the offset by the host galaxy size. We use the host effective radius $R_{\rm eff} = 7.78 \pm 0.03$\,kpc  from morphological fitting \citep{Eftekhari_R155} to calculate a host-normalized offset of $5.13 \pm 0.6~ R_{\rm eff}$, demonstrating a location well outside the locus of the host galaxy light. This large offset of \frb{} from a luminous and quiescent host galaxy makes it stand apart from the majority of the FRB population, characterized primarily by star-forming galaxies at offsets of $\lesssim 10$~kpc \citep{Heintz_2020,Mannings_spiral_gal,Bhandari_2022,Gordon_2023, Sharma2024}. 

For context, in Figure \ref{fig:GC_offsets}, we show the cumulative distribution of projected FRB host galaxy offsets for six repeating and thirty-seven non-repeating FRBs with available offset information in both physical and host-normalized units \citep{Mannings_spiral_gal, Bhandari_2022, Woodland_2024, Sharma2024}. In physical units, the offset of \frb{} exceeds the median FRB population offset of $\approx 5.4$~kpc by a factor of 7; indeed, it has the largest host galaxy offset observed to date. While in host-normalized units, the offset of \frb{} is 4 times larger than the median offset of $\approx 1.4~R_{\rm eff}$, rivaled only by the offset of FRB~20200120E. Many FRB progenitor models invoke a magnetically powered neutron star (NS), called a magnetar, as the central engine \citep{cataloag_of_progenitors_FRBs_Platts}. FRBs localized within star-forming regions can be produced by young magnetars formed via core-collapse supernovae. Thus, while the majority of the FRB population is consistent with magnetars created from core-collapse supernovae, the properties of \frb{} challenge this interpretation, as the fraction of stars at $\sim 5~R_{\rm eff}$ is extremely low. In the following subsections, we briefly consider three progenitor scenarios to explain the large offset from the putative host: a progenitor that (1) formed {\it in situ} in a globular cluster associated with the massive host galaxy, (2) formed {\it in situ} in an undetected satellite dwarf galaxy of the massive galaxy, or (3) was kicked from its birthplace. 

\subsubsection{Globular cluster origin}
For {\it in situ} progenitors, the localization of FRB\,20200120E to a globular cluster \citep[GC;][]{EVN_M81R} demonstrated that at least some FRBs can originate in environments with older stellar progenitors. The extreme offset of \frb{} could naturally be explained by formation in a globular cluster. To place this possibility into context, in Figure~\ref{fig:GC_offsets}, we plot the cumulative distributions of projected physical offsets of GCs in the Milky Way \citep{MW_GC}, M81 \citep{M81_GC_1, M81_GC_2}, and the elliptical galaxies NGC 821 \citep{NGC_821_GC} and NGC 4278 \citep{NGC_4278_GC}. Since the galactocentric offsets of GCs can depend on the sizes of their host galaxies, it is equally informative to compare the host-normalized offsets. Thus, we also show the projected offset distributions in host-normalized units, using $R_{\rm eff}$ of 5.75\,kpc \citep{Reff_MW}, 3.5\,kpc \citep{Reff_M81}, 5.1\,kpc \citep{Reff_NGC_821} and 2.4\,kpc \citep{NGC_4278_GC} for the Milky Way, M81, NGC 821, and NGC 4278, respectively. We also show the projected offset of FRB\,20200120E, which was precisely localized to a GC of the spiral galaxy M81 \citep{Mohit_M81R, EVN_M81R}. We find that in terms of physical offsets, \frb{} is consistent with the upper $\approx 10-25\%$ of the spatial distribution of GCs in ellipticals. When normalized by the host galaxy size, the offset of \frb{} is still consistent with those of GCs in elliptical galaxies in the upper $\approx 30-50\%$ of those distributions. Notably, the host-normalized offset of FRB\,20200120E is very similar to the median value for \frb{} (5.7 vs 5.1~$R_{\rm eff}$). Thus, this comparison clearly illustrates that based on offsets alone, it is plausible that \frb{} could have originated from a globular cluster.

In such a case, promising progenitor models for GC FRBs include magnetars formed via accretion-induced collapse of a white dwarf (WD), or merger induced collapse of a WD-WD, NS-WD, and NS-NS binary, in which dynamical interactions within the GC can increase the rates of such events (e.g., \citealt{Kremer_2021}). Low-mass X-ray binaries (LMXBs) and ultraluminous X-ray sources (ULXs) are also viable given their occurrence in GCs \citep{Clark1975,Maccarone2007,Dage2021, ULX_Sridhar}. Notably, based on deep X-ray measurements, \cite{Aaron_M81R} ruled out ULXs and $\sim30$\% of the brightest Milky Way-like LMXBs for FRB\,20200120E. Thus, the association of \frb{} to a ULX would make it distinct from FRB\,20200120E. 

\subsubsection{Undetected satellite dwarf galaxy}
Another possibility is that the FRB originated in an undetected satellite dwarf galaxy. In such a case, there is no constraint on the age of the progenitor and prompt formation channels such as a magnetar formed via core-collapse supernova of a massive star are possible. However, as discussed in Section \ref{sec:hostgalaxy}, such a galaxy would have $L \lesssim 1.2 \times 10^{7}\,L_{\odot}$ making it $\gtrsim 10$ times less luminous than any other FRB host galaxy with known redshift. Moreover, this upper limit on luminosity falls towards the lower end of the luminosity range possible for the dwarf host of FRB\,20190208A \citep{Dante_dwarfhost}, indicating further that the presence of an undetected satellite dwarf would be an extreme scenario in terms of host luminosity.

In Section \ref{sec:hostgalaxy}, we derived a conservative 95\% upper limit on the redshift of the FRB of $z_\mathrm{max} = 0.19$ assuming 0\,pc\,cm$^{-3}$ for DM$_{\mathrm{MW, halo}}$ and DM$_{\mathrm{host}}$. If we instead use DM$_{\mathrm{MW, halo}}$ = 30\,pc\,cm$^{-3}$ and DM$_{\mathrm{host}}$ = 10\,pc\,cm$^{-3}$, which are still on the lower end of the expected contributions \citep{DM_halo_prochaska_zheng, DM_halo_yamasaki_totani, DM_halo_cook, DM_host_Shin}, we find $z_\mathrm{max} = 0.14$. This suggests that at the redshift of the FRB host galaxy of $z = 0.1384$, there is barely any DM contribution from the local environment and host galaxy of the FRB. This is in stark contrast to the hundreds of units of DM$_{\mathrm{host}}$ for the three FRBs known to reside in dwarf galaxies \citep{R1_host, R1_twin_PRS, Bhandari_2023}. Thus, we find a GC origin to be a more favorable scenario than a satellite dwarf galaxy origin. Space-based imaging coupled with a more precise localization of \frb{} are required to distinguish between the two possibilities. 

\subsubsection{Kicked progenitor}
It is possible that the progenitor of \frb{} was kicked out of its host galaxy. Plausible progenitors include binary neutron star mergers (NS-NS) or WD-NS mergers which in a small fraction of cases may produce indefinitely stable neutron stars \citep{BNS_remnant}. Indeed, the high systemic velocities coupled with the long delay times relative to star formation can result in galactocentric offsets of tens of kpc \citep{BNS_offset}. However, we find that only $\approx 15\%$ ($\approx 5\%$) of short gamma-ray bursts, which likely originate from NS-NS mergers, have physical (host-normalized) offsets comparable to or larger than the offset of \frb{} \citep{Fong_2022}. Furthermore, only $\approx 10\%$ of short gamma-ray burst host galaxies are quiescent \citep{Nugent_2022}. Thus, while the environmental properties of \frb{} are consistent with a subset of kicked compact object binary progenitors, we find it to be a less plausible explanation.

\section{Summary and Conclusion}
The repeating \frb{} was discovered by CHIME/FRB in February 2024, with 22 repeat bursts detected up to July 31st 2024. The high declination of this repeater of $\sim86^{\circ}$ allows a large CHIME exposure time towards the direction of the source, and puts strong constraints on the burst activity before its first detection. We estimated a peak burst rate of $<$20\,hr$^{-1}$ above a fluence threshold of 0.9\,Jy\,ms, almost $10^{4}$ times more than its initial upper limit. This indicates that \frb{} also undergoes sudden episodes of increased activity, as has been seen in other repeaters. Additionally, the burst morphology of \frb{} bursts is consistent with other repeaters. 

The CHIME--KKO baseline provides a 1-D VLBI localization accuracy of $\sim 2^{\prime\prime}$ along the baseline vector. For the six repeat bursts recorded at KKO, the high declination of \frb{} allowed the bursts to be detected with a rotating range of baseline vectors which constrained the combined localization region along several axes. This combined localization was inflated to account for correlated systematic errors, giving a final localization ellipse of dimensions $\sim 1^{\prime\prime}\times2^{\prime\prime}$. 

We observed the field of \frb{} with the Gemini-North telescope and obtained an image of depth $r \gtrsim 25.9$~mag. We use the VLBI localization and PATH to robustly associate \frb{} with a luminous and quiescent elliptical galaxy at $z = 0.1384$ ($P(O|x) = 0.99$). This is the first association of a repeating FRB to a quiescent galaxy, and the first association of any FRB to an elliptical galaxy (see also \cite{Eftekhari_R155}). Moreover, the FRB has a projected physical offset of 40 $\pm$ 5\,kpc from the center of the host galaxy, making it the most offset FRB host to date.  

Given the large offset of the FRB from the host galaxy, we consider a progenitor that is either formed {\it in situ} or kicked from its birthplace. For the {\it in situ} case, an origin in a globular cluster is possible with the projected offset of \frb{} being consistent with the projected offsets of GCs in elliptical galaxies. Moreover, when normalized by the host galaxy size, the offset of \frb{} is comparable to that of FRB~20200120E, providing further support for a GC origin. In such a scenario, promising progenitor models include magnetars formed via accretion induced collapse of a WD, merger-induced collapse of a WD-WD, NS-WD, or NS-NS binary, LMXBs, or ULXs. An extreme scenario would be {\it in situ} formation in an undetected satellite dwarf galaxy of the putative host which is $\gtrsim 10$ fainter than any other FRB host with known redshift. However, DM budgeting barely allows any DM from the host galaxy of the FRB at the redshift of the putative host, making a GC origin more favorable. For a kicked progenitor case, NS-NS or NS-WD mergers which produce stable neutron star remnants are a possibility, where the high kick velocities of these systems can explain the large offset. However, only a small fraction of short gamma-ray bursts, which are thought to be formed from kicked NS-NS mergers, are known to have similar or larger galactocentric offsets, or quiescent host galaxies. Thus, we consider {\it in situ} formation in a GC to be the more likely scenario. 

We thus conclude that the unique local environment and host galaxy of \frb{} revealed by its precise localization adds to the diversity of environments in which repeating FRBs are found. The precise localization of FRBs is crucial to understanding their origins. The CHIME--KKO localization of this repeater provides a proof of concept for the full three-station CHIME/FRB Outrigger array which will be operational in the near future and which will enable sub-arcsecond localizations for thousands of FRBs to their local environments and host galaxies. 

\vspace{2cm}
\section*{ACKNOWLEDGEMENTS}

We acknowledge that CHIME and the \kkoname{} Outrigger (KKO) are built on the traditional, ancestral, and unceded territory of the Syilx Okanagan people. \kkonamecaps{} is situated on land leased from the Imperial Metals Corporation. We are grateful to the staff of the Dominion Radio Astrophysical Observatory, which is operated by the National Research Council of Canada. CHIME operations are funded by a grant from the NSERC Alliance Program and by support from McGill University, University of British Columbia, and University of Toronto. CHIME/FRB Outriggers are funded by a grant from the Gordon \& Betty Moore Foundation. We are grateful to Robert Kirshner for early support and encouragement of the CHIME/FRB Outriggers Project, and to Dusan Pejakovic of the Moore Foundation for continued support. CHIME was funded by a grant from the Canada Foundation for Innovation (CFI) 2012 Leading Edge Fund (Project 31170) and by contributions from the provinces of British Columbia, Québec and Ontario. The CHIME/FRB Project was funded by a grant from the CFI 2015 Innovation Fund (Project 33213) and by contributions from the provinces of British Columbia and Québec, and by the Dunlap Institute for Astronomy and Astrophysics at the University of Toronto. Additional support was provided by the Canadian Institute for Advanced Research (CIFAR), the Trottier Space Institute at McGill University, and the University of British Columbia. The CHIME/FRB baseband recording system is funded in part by a CFI John R. Evans Leaders Fund award to IHS.

We are grateful to Jennifer Andrews and the Gemini Observatory staff for executing our Gemini observations. V.S. is supported by a Fonds de Recherche du Quebec - Nature et Technologies~(FRQNT) Doctoral Research Award. K.S. is supported by the NSF Graduate Research Fellowship Program. C. L. is supported by NASA through the NASA Hubble Fellowship grant HST-HF2-51536.001-A awarded by the Space Telescope Science Institute, which is operated by the Association of Universities for Research in Astronomy, Inc., under NASA contract NAS5-26555. W.F. gratefully acknowledges support by the National Science Foundation under grant no. AST-2206494 and CAREER grant No. AST-2047919, the David and Lucile Packard Foundation, the Alfred P. Sloan Foundation, and the Research Corporation for Science Advancement through Cottrell Scholar Award \#28284. T.E. is supported by NASA through the NASA Hubble Fellowship grant HST-HF2-51504.001-A awarded by the Space Telescope Science Institute, which is operated by the Association of Universities for Research in Astronomy, Inc., for NASA, under contract NAS5-26555. 

B.C.A. is supported by a FRQNT Doctoral Research Award. M.B is a McWilliams fellow and an International Astronomical Union Gruber fellow. M.B. also receives support from the McWilliams seed grant. A.P.C is a Vanier Canada Graduate Scholar. M.D. is supported by a CRC Chair, NSERC Discovery Grant, CIFAR, and by the FRQNT Centre de Recherche en Astrophysique du Qu\'ebec (CRAQ). Y.D. is supported by the National Science Foundation Graduate Research Fellowship under grant No. DGE-2234667. F.A.D is supported by the UBC Four Year Fellowship. E.F. and S.S.P. are supported by the National Science Foundation (NSF) grant AST-2407399. J.W.T.H. and the AstroFlash research group acknowledge support from a Canada Excellence Research Chair in Transient Astrophysics (CERC-2022-00009); the European Research Council (ERC) under the European Union’s Horizon 2020 research and innovation programme (`EuroFlash'; Grant agreement No. 101098079); and an NWO-Vici grant (`AstroFlash'; VI.C.192.045). V.M.K. holds the Lorne Trottier Chair in Astrophysics \& Cosmology, a Distinguished James McGill Professorship, and receives support from an NSERC Discovery grant (RGPIN 228738-13). K.W.M. holds the Adam J. Burgasser Chair in Astrophysics. K.N. is an MIT Kavli Fellow. A.P. is funded by the NSERC Canada Graduate Scholarships - Doctoral program. A.B.P. is a Banting Fellow, a McGill Space Institute~(MSI) Fellow, and a FRQNT postdoctoral fellow. Z.P. is supported by an NWO Veni fellowship (VI.Veni.222.295). M.W.S. acknowledges support from the Trottier Space Institute Fellowship program. P.S. acknowledges the support of an NSERC Discovery Grant (RGPIN-2024-06266). FRB research at UBC is supported by an NSERC Discovery Grant and by the Canadian Institute for Advanced Research. 

Based on observations obtained at the international Gemini Observatory (Program ID: GN-2024A-LP-110), a program of NOIRLab, which is managed by the Association of Universities for Research in Astronomy (AURA) under a cooperative agreement with the National Science Foundation on behalf of the Gemini Observatory partnership: the National Science Foundation (United States), National Research Council (Canada), Agencia Nacional de Investigaci\'{o}n y Desarrollo (Chile), Ministerio de Ciencia, Tecnolog\'{i}a e Innovaci\'{o}n (Argentina), Minist\'{e}rio da Ci\^{e}ncia, Tecnologia, Inova\c{c}\~{o}es e Comunica\c{c}\~{o}es (Brazil), and Korea Astronomy and Space Science Institute (Republic of Korea).

This research has made use of the CIRADA cutout service at URL cutouts.cirada.ca, operated by the Canadian Initiative for Radio Astronomy Data Analysis (CIRADA). CIRADA is funded by a grant from the Canada Foundation for Innovation 2017 Innovation Fund (Project 35999), as well as by the Provinces of Ontario, British Columbia, Alberta, Manitoba and Quebec, in collaboration with the National Research Council of Canada, the US National Radio Astronomy Observatory and Australia’s Commonwealth Scientific and Industrial Research Organisation.

\vspace{5mm}
\facilities{CHIME, KKO, Gemini
(GMOS)}

\software{
\texttt{Astropy} \citep{astropy:2013, astropy:2018, astropy:2022},
\texttt{DM Phase} \citep{dm_phase},
\texttt{fitburst} \citep{fitburst},
\texttt{hdf5} \citep{hdf5},
\texttt{matplotlib} \citep{matplotlib},
\texttt{numpy} \citep{numpy},
\texttt{PyFX} \citep{PyFX}
}

\appendix
\section{Estimating the burst properties} \label{sec:Appendix_burst_properties}

For the bursts with baseband data, the data were beamformed to the best VLBI position of \frb{} listed in Table \ref{tab:VLBI_pos}. The baseband morphology pipeline uses the intensity (Stokes~I) data in the beamformed file to fit a model with burst width, scattering, bandwidth, arrival time, and DM as parameters. While the baseband data have a time resolution of 2.56\,$\mu$s, the data are downsampled in time to achieve a higher S/N per sample for fitting. The pipeline has three main steps, described by \citet{R3_morphology, Sand_2024}. The first step involves radio frequency interference (RFI) excision \citep{baseband_pipeline}, obtaining the structure-maximizing DM \citep{Hessels_structmaxDM} and dedispersion. The second step involves smoothing of the burst profile and estimating the number of burst components. Then, the 1D time series data and spectrum are fit to exponentially modified Gaussians \citep{McKinnon_2014} and running power-law \citep{CHIME_catalog_1} models, respectively, where a Markov Chain Monte Carlo sampling algorithm is used to generate initial parameters for the next step. In the third step, the initial parameters are passed to \texttt{fitburst} \citep{fitburst} to perform a 2D least-square optimization fit on the dynamic spectrum of bursts and return the DM, time of arrival (TOA), signal amplitude, temporal width, power-law spectral index, spectral running and scattering timescale ($\tau$) of the bursts. The scattering time is assumed to go as $\tau \propto \nu^{-4}$ with frequency $\nu$ and with a reference frequency of 600 MHz. The bursts properties for the bursts with intensity data were also estimated using \texttt{fitburst} and a similar procedure as described above. The only difference is the method to estimate the initial guesses for the fit parameters which is described in \citet{CHIME_catalog_1}. 

The fluxes and fluences for the baseband bursts were estimated using the procedure described in \cite{bascat1} and the VLBI position of \frb. The flux calibration pipeline for intensity bursts is set up to use the real-time localization of the burst estimated from the metadata stored during burst detection. Since the uncertainty on these ``header'' localizations can span several degrees, the pipeline can only estimate a lower limit on the flux and fluence \citep{CHIME_catalog_1, Andersen_flux_pipeline}. We thus do not quote flux and fluence measurements for intensity bursts.

Table \ref{tab:burst_prop} lists the TOA, structure-maximizing DM, temporal width, and scattering time returned by \texttt{fitburst} for each of the ten baseband bursts and twelve intensity bursts. The temporal width is the width of the FWHM of the burst envelope at 600\,MHz. Since none of the bursts showed any obvious scattering visually, the scattering times are reported as upper limits, with the upper limit being the width of the narrowest sub-burst. Fluxes and fluences are listed for the ten baseband bursts. Polarization analysis of the \frb{} bursts will be presented in future works.

\begin{deluxetable}{ c c c c c c c c c c }
\tabletypesize{\small}
\tablewidth{0pt} 
\tablecaption{Properties of the repeat bursts from \frb{}.\label{tab:burst_prop}}
\tablehead{
\colhead{Burst} & \colhead{TNS Name} & \colhead{TOA} & \colhead{S/N} & \colhead{DM} & \colhead{Width} & \colhead{$\tau$} & \colhead{t$_{\text{res}}$} & \colhead{Peak Flux} & \colhead{Fluence}   \\
\colhead{} & \colhead{} & \colhead{} & \colhead{} & \colhead{(pc cm$^{-3}$)}& \colhead{(ms)} & \colhead{(ms)} & \colhead{(ms)} & \colhead{(Jy)} & \colhead{(Jy ms)}}
 \startdata
B1$^*$ & {FRB 20240209A} & 2024-02-09 07:10:14 & 15.99 & 176.49$\pm$0.01 & 23.4 & $<$0.2 & 0.082 & 20.9$\pm$2.5& 324.9$\pm$33.4\\
B2 & {FRB 20240217A} & 2024-02-17 06:36:05 & 9.24 & 177.63$\pm$0.24 & 29.7 & $<$4.58 & 0.983 & $\mathord{\cdot}\mathord{\cdot}\mathord{\cdot}$ & $\mathord{\cdot}\mathord{\cdot}\mathord{\cdot}$ \\
B3$^*$ & {FRB 20240309A} & 2024-03-09 04:29:37 & 16.78 & 176.33$\pm$0.03 & 1.9 & $<$0.15 & 0.041 & 7.0$\pm$0.8 & 12.9$\pm$1.5\\
B4 & {FRB 20240608A} & 2024-06-08 10:23:09 & 9.0 & 175.4$\pm$0.09 & 49.9 & $<$2.99 & 0.983 & $\mathord{\cdot}\mathord{\cdot}\mathord{\cdot}$ & $\mathord{\cdot}\mathord{\cdot}\mathord{\cdot}$ \\
B5 & {FRB 20240612A} & 2024-06-12 08:51:06 & 14.42 & 175.07$\pm$0.14 & 32.1 & $<$0.79 & 0.655 & 2.9$\pm$0.6 & 18.2$\pm$2.4 \\
B6 & {FRB 20240612B} & 2024-06-12 21:33:20 & 10.89 & 176.82$\pm$0.51 & 45.5 & $<$19.32 & 0.983 & $\mathord{\cdot}\mathord{\cdot}\mathord{\cdot}$ & $\mathord{\cdot}\mathord{\cdot}\mathord{\cdot}$ \\
B7 & {FRB 20240616A} & 2024-06-16 22:52:19 & 13.7 & 175.27$\pm$0.04 & 3.8 & $<$1.09 & 0.164 & 18.7$\pm$2.7 & 64.9$\pm$8.0 \\
B8 & {FRB 20240619A} & 2024-06-19 08:57:36 & 10.46 & 175.12$\pm$0.01 & 57.4 & $<$0.72 & 0.328 & 3.9$\pm$0.8 & 21.4$\pm$2.6 \\
B9 & {FRB 20240619B} & 2024-06-19 09:03:37 & 10.73 & 175.19$\pm$0.02 & 17.7 & $<$0.22 & 0.328 & 2.8$\pm$0.5 & 21.2$\pm$2.4 \\
B10 & {FRB 20240619C} & 2024-06-19 10:11:00 & 10.3 & 176.51$\pm$0.18 & 16.2 & $<$6.88 & 0.983 & $\mathord{\cdot}\mathord{\cdot}\mathord{\cdot}$ & $\mathord{\cdot}\mathord{\cdot}\mathord{\cdot}$ \\
B11 & {FRB 20240620A} & 2024-06-20 08:55:51 & 8.47 & 175.35$\pm$0.14 & 55.5 & $<$1.91 & 0.983 & $\mathord{\cdot}\mathord{\cdot}\mathord{\cdot}$ & $\mathord{\cdot}\mathord{\cdot}\mathord{\cdot}$ \\
B12 & {FRB 20240620B} & 2024-06-20 09:14:46 & 8.53 & 175.91$\pm$0.27 & 33.2 & $<$4.02 & 0.983 & $\mathord{\cdot}\mathord{\cdot}\mathord{\cdot}$ & $\mathord{\cdot}\mathord{\cdot}\mathord{\cdot}$ \\
B13 & {FRB 20240621A} & 2024-06-21 21:00:56 & 9.02 & 179.46$\pm$0.77 & 63.2 & $<$26.85 & 0.983 & $\mathord{\cdot}\mathord{\cdot}\mathord{\cdot}$ & $\mathord{\cdot}\mathord{\cdot}\mathord{\cdot}$ \\
B14 & {FRB 20240625A} & 2024-06-25 20:20:30 & 12.29 & 175.39$\pm$0.04 & 41.7 & $<$1.27 & 0.983 & $\mathord{\cdot}\mathord{\cdot}\mathord{\cdot}$ & $\mathord{\cdot}\mathord{\cdot}\mathord{\cdot}$ \\
B15 & {FRB 20240628A} & 2024-06-28 22:05:21 & 9.59 & 175.72$\pm$0.16 & 30.5 & $<$5.83 & 0.983 & $\mathord{\cdot}\mathord{\cdot}\mathord{\cdot}$ & $\mathord{\cdot}\mathord{\cdot}\mathord{\cdot}$ \\
B16$^*$ & {FRB 20240629A} & 2024-06-29 07:58:35 & 16.17 & 175.18$\pm$0.01 & 16.4 & $<$0.24 & 0.082 & 9.6$\pm$1.1 & 113.7$\pm$11.6 \\
B17$^*$ & {FRB 20240629B} & 2024-06-29 08:32:47 & 20.67 & 175.17$\pm$0.01 & 15.6 & $<$0.18 & 0.041 & 12.5$\pm$1.4 & 46.5$\pm$5.0\\
B18$^*$ & {FRB 20240629C} & 2024-06-29 08:57:54 & 19.46 & 175.21$\pm$0.04 & 2.7 & $<$1.13 & 0.655 &  4.6$\pm$0.8 & 7.4$\pm$1.3 \\
B19 & {FRB 20240629D} & 2024-06-29 11:07:02 & 9.7 & 176.47$\pm$0.43 & 30.4 & $<$12.93 & 0.983 & $\mathord{\cdot}\mathord{\cdot}\mathord{\cdot}$ & $\mathord{\cdot}\mathord{\cdot}\mathord{\cdot}$ \\
B20$^*$ & {FRB 20240629E} & 2024-06-29 19:16:34 & 16.12 & 175.37$\pm$0.01 & 6.0 & $<$0.44 & 0.164 & 30.1$\pm$5.0 & 97.1$\pm$13.1 \\
B21 & {FRB 20240702A} & 2024-07-02 08:36:06 & 9.25 & 177.29$\pm$0.3 & 20.9 & $<$8.87 & 0.983 & $\mathord{\cdot}\mathord{\cdot}\mathord{\cdot}$ & $\mathord{\cdot}\mathord{\cdot}\mathord{\cdot}$ \\
B22 & {FRB 20240716A} & 2024-07-16 06:14:15 & 8.6 & 175.3$\pm$0.15 & 17.7 & $<$7.5 & 0.983 & $\mathord{\cdot}\mathord{\cdot}\mathord{\cdot}$ & $\mathord{\cdot}\mathord{\cdot}\mathord{\cdot}$ \\
\enddata
\tablecomments{The burst number is used to refer to the individual bursts. $^*$ indicates the bursts detected at KKO and localized with the CHIME--KKO baseline. Although each burst has a different TNS name, all bursts are from the same repeating source \frb{}, which has the TNS name of the first burst from this repeater. Topocentric TOAs are referenced at 400 MHz and calculated using the structure-maximizing DM indicated in the fifth column and a DM constant of $K_{\text{DM}} = 10^4/2.41\,\text{s}\,\text{MHz}^2\,\text{pc}^{-1}\,\text{cm}^3$. The scattering timescale $\tau$ is referenced at 600 MHz and reported as an upper limit equivalent to the width of the narrowest subburst. All burst properties are calculated at a time resolution t$_{\text{res}}$. Fluxes and fluences are not quoted for intensity bursts as our current pipelines can only estimate a lower limt on these values.}
\end{deluxetable}

\section{Systematic error in CHIME--KKO localizations \label{Appendix_B}}
There can be multiple sources of systematic errors in the CHIME--KKO localizations. The residual ionospheric delays are expected to be a source of time-dependent systematic errors that are not correlated for data taken at different times. Another source of systematic error is the beam phase. The primary beam of each feed in the CHIME and KKO stations has an unknown phase that is not accounted for while obtaining geometric delays from the phase calibrated visibilities \citep{KKO_overview}. The beam phase is expected to be time-independent and spatially dependent on the pointing of the correlator on the sky. Due to the high declination of \frb{}, the correlator pointing for each of the six repeat bursts is similar, and thus the systematic errors in the localizations due to beam phase are correlated.  

To obtain the correlated systematic error in the combined CHIME--KKO localization, we used 3 -- 5 localizations from 12 pulsars each, which were obtained using the procedure described in Section \ref{sec:localization} and by \cite{KKO_overview}. For each localization, the calibrator that was closest to the pulsar on the sky and detectable within the same baseband dump as the pulsar pulse was used as the phase calibrator. Only the frequency channels used for \frb{} localizations were used for the pulsar localizations, to remove any bandwidth dependence of errors. The multiple localizations were combined for each pulsar, and the offset of the combined localization position from the true pulsar position along the baseline vector was obtained. These ``projected" offsets had an RMS of $\sim1^{\prime\prime}$. We assume that combining multiple localizations averaged down time-dependent errors, and thus the correlated systematic error in CHIME--KKO localizations is $\sim1^{\prime\prime}$. While the conclusions of \cite{KKO_overview} tell us that the RMS localization error in CHIME--KKO localizations for broadband one-off bursts is $\sim$$1^{\prime\prime}$, the analysis here indicates that the RMS error in combined localizations of narrowband bursts from repeaters is also $\sim$$1^{\prime\prime}$. 

Note that the beam phase error depends on the separation of the target and calibrator on the sky. The pulsar localizations had target-calibrator separations of tens of degrees, while the target-calibrator separation for \frb{} is $\sim$4 degrees. We expect the beam phase error in the \frb{} localizations to be similar for both the target and the calibrator since they are close together on the sky, and to be removed upon calibration. Thus, if the systematic error is dominated by the beam phase error, assuming a systematic error of $1^{\prime\prime}$ in the combined CHIME--KKO localization for \frb{} gives us a conservative localization error. 
\bibliography{FRB20240209A}
\bibliographystyle{aasjournal}
\end{document}